\documentclass[aps,prb,reprint,amsmath,showpacs,superscriptaddress, longbibliography]{revtex4-1}
\usepackage{graphicx}
\usepackage{textcomp}
\usepackage{float}
\usepackage{xcolor}
\usepackage[normalem]{ulem}
\usepackage{upgreek}
\usepackage{amsfonts}

\newcommand{\kl}[1]{\ensuremath{ \left( #1 \right) }}

\usepackage{tikz}
\definecolor{myred}{HTML}{FF1F5B}
\definecolor{myorange}{HTML}{F28522}
\definecolor{mygreen}{HTML}{00CD6C}
\usepackage{braket}
\usepackage{cleveref}
\usepackage{amssymb}

\DeclareMathSymbol{\shortminus}{\mathbin}{AMSa}{"39}

\begin{document}

\title{Mapping quantum Hall edge states in graphene by scanning tunneling microscopy}

\author{T.~Johnsen}
    \affiliation{2nd Institute of Physics B and JARA-FIT, RWTH Aachen University, 52074 Aachen, Germany}
\author{C.~Schattauer}
    \affiliation{Institute for Theoretical Physics, TU Wien, Wiedner Hauptstraße 8-10, 1040 Vienna, Austria}
\author{S.~Samaddar}
   \affiliation{2nd Institute of Physics B and JARA-FIT, RWTH Aachen University, 52074 Aachen, Germany}
    \affiliation{National Physical Laboratory, Teddington, United Kingdom} 
\author{A.~Weston}
    \affiliation{Department of Physics and Astronomy, University of Manchester, Manchester, United Kingdom}
    \affiliation{National Graphene Institute, University of Manchester, Manchester, United Kingdom}
\author{M.~J. Hamer}
    \affiliation{Department of Physics and Astronomy, University of Manchester, Manchester, United Kingdom}
     \affiliation{National Graphene Institute, University of Manchester, Manchester, United Kingdom}
\author{K.~Watanabe}
    \affiliation{National Institute for Materials Science, Tsukuba, Ibaraki 305-0044, Japan}
\author{T.~Taniguchi}
    \affiliation{National Institute for Materials Science, Tsukuba, Ibaraki 305-0044, Japan}
\author{R.~Gorbachev}
    \affiliation{Department of Physics and Astronomy, University of Manchester, Manchester, United Kingdom}
    \affiliation{National Graphene Institute, University of Manchester, Manchester, United Kingdom}
    \affiliation{Henry Royce Institute for Advanced Materials, University of Manchester, Manchester, United Kingdom}
\author{F.~Libisch}
    \affiliation{Institute for Theoretical Physics, TU Wien, Wiedner Hauptstraße 8-10, 1040 Vienna, Austria}
\author{M.~Morgenstern}
     \email{Corresponding author: mmorgens@physik.rwth-aachen.de}
    \affiliation{2nd Institute of Physics B and JARA-FIT, RWTH Aachen University, 52074 Aachen, Germany}

\date{\today}

\begin{abstract}
Quantum Hall edge states are the paradigmatic example of the bulk-boundary correspondence. They are prone to intricate reconstructions calling for their detailed investigation  at high spatial resolution. Here, we map quantum Hall edge states of monolayer graphene at a magnetic field of 7\,T with scanning tunneling microscopy. Our graphene sample features a gate-tunable lateral interface between areas of different filling factor.  We compare the results with detailed tight-binding calculations quantitatively accounting for the perturbation by the tip induced quantum dot. We find that an adequate choice of gate voltage allows for mapping the edge state pattern with little perturbation.
We observe extended compressible regions, the antinodal structure of edge states and their meandering along the lateral interface.
\end{abstract}


\maketitle
\section{Introduction}
\label{sec:intro}
The quantum Hall (QH) effect \cite{Klitzing1980} initiated the topological description of electron systems in solids \cite{Thouless1982,Kohmoto1985,Niu1985}. The principle of bulk-boundary correspondence attributes the bulk related Chern number to edge states carrying the dissipationless Hall current.\cite{Halperin1982,Hatsugai1993,Graf2013} This revolutionary insight triggered more detailed investigations of the spatial structure of the edge states in the presence of Coulomb interactions, starting with calculations of the widths of compressible and incompressible stripes.\cite{Chklovskii1992,Lier1994} Later, complex reconstructions including charged and neutral upstream modes have been predicted at filling factor $\nu=1$ \cite{Khanna2021,Venkatachalam2012,Bhattacharyya2019} and in the fractional QH regime.\cite{MacDonald1990,Kane1994,Wan2003,Wang2013}  The neutral upstream modes have partially been evidenced indirectly via their shot noise properties.\cite{Bid2010,Sabo2017}  For graphene, upstream modes can also appear due to the bare gate electrostatics at the rim of the graphene flake.\cite{Marguerite2019,Moreau2021}
Moreover, a fragmentation of integer QH edge states by exchange interactions has been predicted.\cite{Oswald2021,Oswald2017,Oswald2017b} These intriguing predictions and indirect experimental results call for a more detailed spatial investigation of QH edge states in real space.

Initial studies used a scanning single electron transistor \cite{Yacoby1999} as well as electrostatic force microscopy \cite{McCormick1999,Weitz2000,Weis2011} to evidence the presence of edge states. Later, scanning gate microscopy,\cite{Aoki2005,Paradiso2012,Pascher2014} scanning capacitance microscopy,\cite{Suddards2012}, microwave impedance microscopy,\cite{Lai2011} and scanning SQUID microscopy \cite{Uri2019} have been employed. However, all of these methods provide a spatial resolution well above the magnetic length, such that the internal structure of the edge states remained elusive. Indirectly, macroscopic tunneling experiments
probed the internal structure via the in-plane $B$ field dependence, but only for rather steep potential profiles without signatures of reconstruction and, naturally, without any information on the edge state pattern along the edge.\cite{Patlatiuk2020}
Hence, higher resolution scanning probes are mandatory for this purpose. They can be favorably applied to graphene with its exposed surface.\cite{Andrei2012,Morgenstern2011} Indeed, imaging of quantum Hall edge states by scanning tunneling microscopy (STM) has been attempted  at a graphene boundary, but on a strongly screening graphite substrate that naturally suppresses any edge state evolution.\cite{Li2013a} More recently, a gated lateral interface of graphene on h-BN has been employed to realize more soft confinement,\cite{Kim2021} enabling the visualization of symmetry broken edge states by Kelvin probe force microscopy. Nevertheless, the internal internal edge state structure has remained elusive.

\begin{figure*}
    \centering
    \includegraphics[scale=0.6]{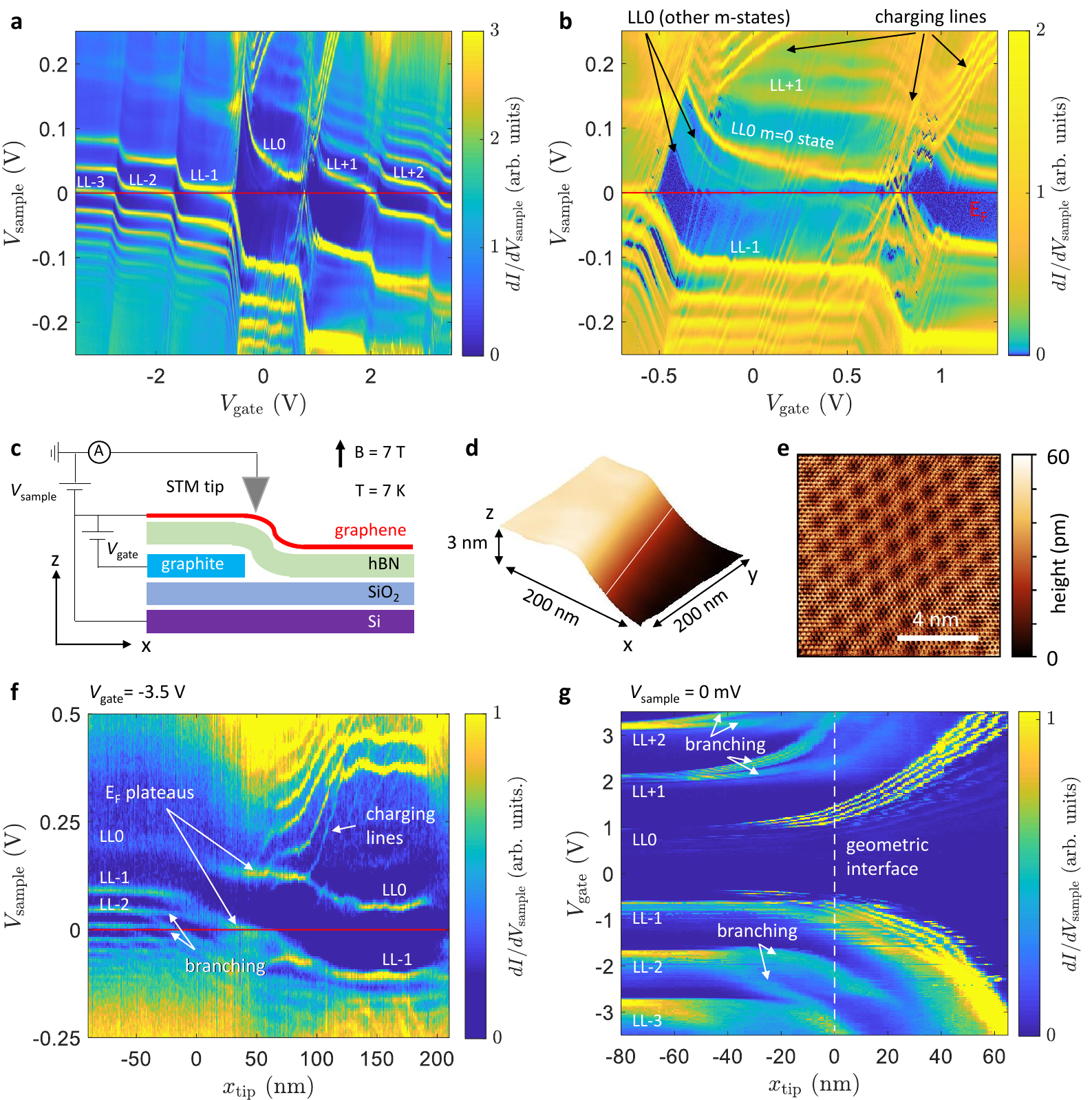}
    \caption{
    {\bf Scanning tunneling spectroscopy of the lateral interface at $B=7$\,T:}
    (a) $dI/dV_{\rm sample}(V_{\rm gate},\,V_{\rm sample})$ on a  graphene area far away from the lateral interface. Landau level features LL$n$ are marked. The tip-sample distance is stabilized at current $I_{\rm stab}=1$\,nA and voltage $V_{\rm stab}=-250$\,mV.
    (b) Zoom into the area where the LL0 lines cross $E_{\rm F}$ ($V_{\rm sample}=0$\,V). The marked bright line above $E_{\rm F}$ corresponds to the  ($m=0$)-orbital of LL0 confined in the TIQD. The replica of this line at lower $V_{\rm sample}$ are other confined $m$-states of LL0. Charging lines run from the lower left to the upper right. The ones that cross an $m$-state at $E_{\rm F}$ are caused by the charging of exactly this $m$-state. Quadruplets of  charging lines showcase the spin and valley degeneracy of graphene. 
    (c) Sample layout with circuitry, graphite thickness: 3\,nm, hBN: 23\,nm, SiO$_2$: 300\,nm. The graphite is used to partially gate the graphene.
    (d) STM topography of graphene with a step marking the onset of the underlying graphite  defined as $x_{\rm tip}=0$\,nm (white line), $I=200$\,pA, $V_{\rm sample}=-500$\,mV.
    (e) STM topography of graphene with atomic resolution and moir\'e lattice due to a mutual rotation of the graphene and the underlying hBN by 11.1$^\circ$, $I=1$\,nA, $V_{\rm sample}=-250$\,mV, $V_{\rm gate}=3.5\,$V.
    (f) $dI/dV_{\rm sample} (x_{\rm tip},\,V_{\rm sample})$ across the lateral interface, $I_{\rm stab}=200$\,pA,  $V_{\rm stab}=-500$\,mV, $V_{\rm gate}=-3.5\,\rm V$.
    (g) $dI/dV_{\rm sample}(x_{\rm tip},\,V_{\rm gate})$ across the lateral interface,  $I_{\rm stab}=1$\,nA, $V_{\rm stab}=-250$\,mV, $V_{\rm sample}=0$\,V.
    }
    \label{Fig1}
\end{figure*}

Here, we apply scanning tunneling microscopy (STM) in a perpendicular magnetic field  $B=7$\,T \cite{Mashoff2009} probing an interface between different filling factors. The well-known Landau level (LL) pinning at the Fermi level $E_{\rm F}$ as function of gate voltage $V_{\rm gate}$
\cite{Jung2011,Chae2012,Walkup2020} causes LL plateaus across the interface indicating the appearance of compressible stripes.\cite{Chklovskii1992,Lier1994}
As a major challenge, the electrostatic potential of the tip itself induces a quantum dot immediately below the tip.\cite{Dombrowski1999,Freitag2016} This tip-induced quantum dot (TIQD) can significantly affect the measurement, locally disturbing the edge-state structure one hopes to measure. Here, we use  experimentally observed charging lines to deduce the parameters determining width and depth of the TIQD in detail.\cite{Freitag2016,Freitag2018} We thereby set up a tight-binding (TB) model that quantitatively accounts for the local electrostatics around the tip, and hence include effects of the TIQD. Comparing the measured $dI/dV$ signal as function of $V_{\rm gate}$, sample voltage $V_{\rm sample}$ and position $x_{\rm tip}$ to our simulations allows us to identify parameter regimes where the perturbation due to the TIQD is minimal, enabling the spatial mapping of the edge states with unprecedented resolution.

In detail, we calculate the local density of states (LDOS) below the tip center as we virtually move the TIQD across the interface at $B=7$\,T. Our model reproduces all features found in the experiment and, hence, enables a direct comparison with the spatial LDOS distribution for each tip position $x_{\rm tip}$. We find that the dominant lines of $dI/dV_{\rm sample}(x_{\rm tip},\,V_{\rm sample}, V_{\rm gate})$  are caused by states of the TIQD. However, weaker branching-type features at the interface represent the barely perturbed edge states with inner anti-nodal structure.\cite{Ando1984,Joynt1984,Bindel2017} Using this novel insight, we measure the spatial distribution of the edge states for various LLs. They show the antinodal structure of LL wave functions that is slightly modified by the potential gradient at the interface and likely also by the electron-electron repulsion. Moreover, the first hole-type LL edge state is mapped along the interface revealing the expected meandering and a spatial variation of its internal structure.

\section{Experimental Setup}
\label{sec:exsetup}
We prepare the graphene sample by the dry stacking method depositing a sequence of 3\,nm thick graphite,  23~nm thick hBN and a monolayer graphene exfoliated from graphite on top of Si/SiO$_2$ (Fig.~\ref{Fig1}c). The graphene is placed partially above the graphite flake to create a tunable potential step (Fig.~\ref{Fig1}c). Graphite and graphene are contacted by Au electrodes via shadow mask evaporation (Appendix Section~\ref{sec:prep_sample}). An STM operating at 7\,K in ultrahigh vacuum up to $B=7$\,T probes the LDOS at varying $V_{\rm gate}$ applied to the graphite, i.e. potential drop across the interface.\cite{Mashoff2009} An additional voltage $V_{\rm sample}$ is applied to the graphene with respect to the grounded tip that records the tunneling current $I$ (Fig.~\ref{Fig1}c).
The $dI/dV_{\rm sample}(V_{\rm sample})$ recorded by lock-in technique is (to first order) proportional to the LDOS at energy $E-{E_{\rm F}}=eV_{\rm sample}$.\cite{Morgenstern2000b} An additional numerical derivative $d^2I/dV_{\rm sample}dV_{\rm gate}$ improves the visibility of the charging lines.
\section{Computational Details}
\label{sec:CompDetails}
For the TB calculations, we use a $3^{\rm rd}$ nearest-neighbour hopping model \cite{PhysRevB.102.155430} for a rectangular single layer  graphene flake (220\,nm$\times$400\,nm) reading
\begin{equation}
\mathcal{H} = \sum_i s_i \hat{c}^{\dagger}_i \hat{c}_i +
\sum_{\langle i,j\rangle} \gamma_{ij} \mathrm{e}^{2\pi\text{i}\delta_{ij}}\hat{c}^{\dagger}_i \hat{c}_j,
\label{eq:H_TB2}
\end{equation}
where $s_i$ is the on-site energy at site $i$, $\gamma_{ij}$ are hopping parameters between site $i$ and site $j$, $\hat{c}^{\dagger}_i$ ( $\hat{c}_i$) are creation (annihilation) operators at site $i$, and the $B$ field is included via a Peierls phase
\begin{equation}
\delta_{ij}=\frac{1}{\Phi_0}\int_{\mathbf{r}_i}^{\mathbf{r}_j} \mathbf{A} \cdot \mathrm{d}\mathbf{r}
\label{eq:Peierl}
\end{equation}
with the magnetic flux $\Phi_0=h/e$, 
positions $\mathbf{r}_i$, ${\mathbf{r}_j}$ of sites $i$, $j$,
and the vector potential in Landau gauge $\mathbf{A}=B x \mathbf{\hat{y}}$.
To simulate the large dimensions of the experimental flake, we employ a rescaled graphene Hamiltonian.\cite{Liu2015} It increases the interatomic distances by a factor of ten and accordingly reduces the dimension of the Hamiltonian without qualitatively altering the energy spectra.
This approximation holds since we do not expect the lattice scale of graphene to be relevant at the large scale of the TIQD ($10-100$\,nm) and the magnetic length ($\sim 10$\,nm).\cite{Liu2015}

\section{Results}
\label{sec:results}
\subsection{Spectroscopy at a Single Location}
\label{sec:singlespot}
Figure~\ref{Fig1}a--b
shows $dI/dV_{\rm sample}(V_{\rm gate},\,V_{\rm sample})$ recorded at a tip location $x_{\rm tip}$ above the graphite gate far away from the interface ($B=7$\,T). States belonging to the Landau levels LL$n$ ($n\in \mathbb{Z}$) are visible as bright LDOS lines that feature steps in the $(V_{\rm gate},V_{\rm sample})$ plane caused by the pinning of LLs to $E_{\rm F}$.\cite{Jung2011,Luican2011,Chae2012} The LL indices $n$ are identified by the mutual energy distance of the lines and are accordingly marked.
Plateaus appear close to $E_{\rm F}$ at hole doping ($V_{\rm gate}< -1$\,V), but show up at an energy distinct from $E_{\rm F}$ at electron doping and for LL0.\cite{Jung2011} This implies that the measured LDOS lines are not caused by the intrinsic LLs of the unperturbed graphene bulk, but rather by states of the TIQD.\cite{Dombrowski1999} The confined states of a quantum dot in $B$ field are roughly classified for each LL$n$ by their different azimuthal quantum numbers $m$ (Appendix Section~\ref{sec:TB_TIQD}).\cite{Schnez2008,Freitag2016} The most prominent LDOS line belongs to the ($m=0$)-state \cite{Morgenstern2000,Freitag2016} as the only one with an antinode in the TIQD center,\cite{Schnez2008} where the tip is probing $I$ (Appendix Section~\ref{sec:TB_TIQD}). Additional weaker lines in $dI/dV_{\rm sample}(V_{\rm gate},\,V_{\rm sample})$ run in parallel to the ($m=0$)-state of LL0 at lower $V_{\rm sample}$ (Fig.~\ref{Fig1}b, zoom-in at larger contrast in Fig.~\ref{FigS6a}). They correspond to ($|m|>0$)-states with higher confinement energy.\cite{Schnez2008} The fact that the ($|m|>0$)-states are below the ($m=0$)-state classifies the TIQD as hole-type (Appendix Sections~\ref{sec:ParametersPoisson}-\ref{sec:TB_TIQD}). The apparent plateaus of the $m$-states are eventually caused by the pinning of the bulk LLs of graphene at $E_{\rm F}$. The pinning prohibits a strong change of the TIQD depth by $V_{\rm gate}$. Hence, the plateaus at $E_{\rm F}$ for $V_{\rm gate} < -1$\,V imply that the ($m=0$)-state of the TIQD is barely displaced energetically from the corresponding bulk LL$n$, i.e. the depth of the TIQD is shallow.

Besides the LDOS lines belonging to $m$-states of different LL$n$, the measured $dI/dV$ curves (Fig.~\ref{Fig1}a--b) feature additional lines that are tilted oppositely to the LDOS lines (lines marked "charging lines" in Fig.~\ref{Fig1}b) . They are charging lines \cite{Jung2011,Freitag2016,Chae2012,Walkup2020} with a slope caused by the positive gate voltage compensating a negative tip voltage (hence a positive $V_{\rm sample}$) to keep the charge in the TIQD constant. Such charging lines are known to be caused by the Coulomb staircase effect,\cite{Freitag2016, Freitag2018} i.e., each additional electron added to the TIQD changes the LDOS and thus the measured current abruptly by Coulomb repulsion. The jumps in the LDOS associated with these charging events thus appear prominently in scanning tunneling spectroscopy (Fig.~\ref{Fig1}b), e.g., the ($m=0$)-states of LL0 and LL-1 exhibit kinks wherever a charging line crosses (see also Fig.~\ref{FigS6a}, appendix).  
Some charging lines exhibit quadruplets with regular distances as expected from the fourfold degeneracy of graphene  (Appendix Section~\ref{sec:fixedlateral}).\cite{Freitag2016}

The most prominent charging lines cross the LDOS features marked LL$n$ at the right end of their plateaus at $E_{\rm F}$ (Fig.~\ref{Fig1}a, b, Fig.~\ref{FigS6b}, appendix). Weaker charging lines follow towards the left (i.e., for smaller $V_{\mathrm{gate}}$. This again classifies the TIQD as hole-type,\cite{Jung2011} since the ($m=0$)-state is charged firstly with highest impact on the probed LDOS due to its antinode directly below the tip (Appendix Section~\ref{sec:ParametersPoisson}). A more negative $V_{\rm gate}$ removes further electrons, i.e. charges holes into higher $m$-states of the TIQD, that feature a larger average lateral distance to the tip center and, hence, induce less changes of the LDOS below the tip. We use the charging lines to determine energetic depth and lateral width of the TIQD below.
\begin{figure*}
    \centering
    \includegraphics[width=0.999\textwidth]{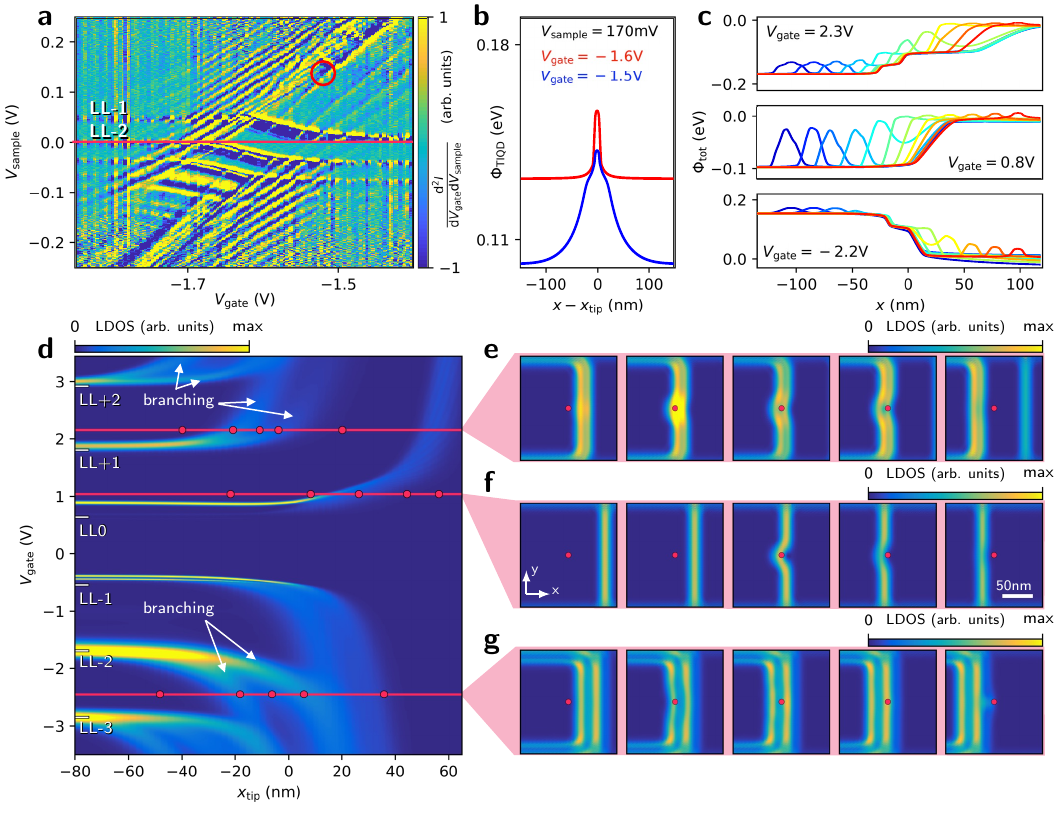}
    \caption{
        {\bf Origin of the branching of LL$n$ states at the interface:}
        (a) $d^2I/dV_{\rm sample}dV_{\rm gate}(V_{\rm gate},\,V_{\rm sample})$ at the transition from LL-2  to LL-1 being located at $E_{\rm F}$ (red line), $I_{\rm stab}=1$\,nA, $V_{\rm stab}=-250$\,mV. The crossing point of the first ($m=0$)-charging line of LL-2  with the last charging line belonging to LL-1 is marked (red circle). Such crossings for various adjacent LL$n$ are used to determine $\Delta V_{\rm sample}$ and $\Delta V_{\rm gate}$ as input parameters for the Poisson calculations (Appendix Section~\ref{sec:ParametersPoisson}).
        (b) Potential of the TIQD without lateral interface for the marked $V_{\rm sample}$, $V_{\rm gate}$ as resulting from the Poisson simulations.
        (c) Profile line through the potential of superposed TIQD and lateral interface,
        $V_{\rm sample}=0$\,V, $x_{\rm tip} \in [-130,+100]$\,nm with increments of 20\,nm (blue to red), $V_{\rm gate}$ as marked. The potentials are used as input for the TB simulations.
        (d) LDOS($x_{\mathrm{tip}}$,\,$V_{\mathrm{gate}}$) at $V_{\rm sample}=0$\,V as resulting from the TB simulations (Appendix Sections~\ref{sec:TB_TIQD}--\ref{sec:InterpolTB}).
        The LDOS is averaged over a circular region (radius $\approx 1.5$ nm) around the vector $\mathbf{x}_{\mathrm{tip}}$ describing the position of the tip center. White horizontal lines on the left mark the bulk LL$n$. The marked branching of various LL$n$ states around $x_{\rm tip}=0$\,nm qualitatively matches the experimental ones (Fig.~\ref{Fig1}g). Red lines with red dots mark $V_{\rm gate}$ and $\mathbf{x}_{\rm tip}$, respectively, as used in e--g.
        (e) LDOS as a function of real space coordinates $x,y$, $V_{\mathrm{gate}}= 2.15$\,V. The columns are for different $\mathbf{x}_{\mathrm{tip}}$  marked by red dots (also in d). (f) Same as e,  $V_{\mathrm{gate}}= 1.05$\,V. (g) Same as e, $V_{\rm gate}=-2.45$\,V.
        }
    \label{Fig2}
\end{figure*}

\subsection{Spectroscopy across the Interface}
\label{sec:specInterface}
The sample geometry (Fig.~\ref{Fig1}c) enables a lateral interface of different filling factors via a partial graphite gate that changes the carrier density in  the graphene area on the left.
The position of the interface is determined by STM as a visible step of the graphite height within the graphene layer (Fig.~\ref{Fig1}d). The graphene is strongly rotated ($11.1^\circ$) with respect to the underlying hBN (Fig.~\ref{Fig1}e) minimizing the influence of the moir\'e structure on the TIQD states.\cite{Freitag2016,Freitag2018}

Moving the STM tip across the interface while varying $V_{\rm sample}$ or $V_{\rm gate}$ reveals the evolution of the Landau level energies across the interface (Fig.~\ref{Fig1}f and g) . On the left (right) in Fig.~\ref{Fig1}f, $E_{\rm F}$ is between LL-4 and LL-3 (LL-1 and LL0).
Hence, filling factors are different as intended.  The LL-1 state exhibits a plateau at $E_{\rm F}$ close to the interface ($x_{\rm tip}=0$\,nm). This demonstrates a lateral pinning of the bulk LL-1 to $E_{\rm F}$ typically dubbed a compressible stripe \cite{Chklovskii1992} and a rather shallow TIQD. For the unoccupied LL0 state, we observe a similar plateau, but slightly shifted to the right (arrows labeled "$E_F$ plateaus" in Fig.~\ref{Fig1}f). This confirms the flat potential area caused by the compressible stripe and demonstrates a change of the compressible stripe position by the  TIQD, respectively.  Almost horizontal charging lines appear in the upper right of Fig.~\ref{Fig1}f highlighting the charging of LL-1 states that are pulled across $E_{\rm F}$ by positive $V_{\rm sample}$.\cite{Hashimoto2008} The $V_{\rm sample}$ evolution of these charging lines along $x_{\rm tip}$ directly probes the potential \cite{Hashimoto2008} and hence again confirms the rather flat potential areas in the interface region ($x_{\rm tip}\simeq 50$\,nm) as expected for compressible stripes  as well as surrounding steeper potentials featuring the separating incompressible stripes.\cite{Chklovskii1992,Lier1994}
The three charging lines that subsequently propagate along the LDOS plateau of LL0  (label "charging lines" in Fig.~\ref{Fig1}f) showcase, moreover, the charge carrier density gradient at constant potential as expected for compressible stripes.\cite{Chklovskii1992}
Finally, we find an unexpected branching of the LL-2 and LL-1 line around $x_{\rm tip} \in[-50,0]$\,nm that we analyze in detail below. Notice that a branching of LLs has also been observed for topological insulator states within a remote Coulomb potential, where $V_{\rm gate}$ could not be applied.\cite{Fu2014} It has been attributed to a similar origin as in our analysis, i.e. to the antinodal structure of the LL wave functions.\cite{Fu2014}

We next consider the evolution of $dI/dV_{\rm sample}$ at $E_{\rm F}$ ($V_{\rm sample}=0\,$V) across the interface for varying height of the potential step by tuning $V_{\mathrm{gate}}$ (Fig.~\ref{Fig1}g). The strongest $V_{\rm gate}$ influence is observed on the left side ($x_{\mathrm{tip}} < 0$) as expected, where many LL$n$ cross $E_{\rm F}$. However, the gate continues to influence LL$n$ features, at least, up to $x_{\mathrm{tip}} = 60\,$nm, to the right of the graphite gate, albeit to a weaker extent. Quadruplets of lines are partially observed, e.g. in the upper right, 
implying a strong influence of the charging of the TIQD at the corresponding voltages. More interestingly, a pronounced branching of the LL$n$ states across the interface appears again for LL+2, LL+1, and LL-2 (labels "branching" in Fig.~\ref{Fig1}g). The branching is barely visible for LL-1 exhibiting only a shoulder at the left of the main intensity. This is due to the interference of charging lines. But the branching of LL-1 is clearly apparent in two-dimensional maps of this LL at $E_{\rm F}$, where a double line is meandering along the interface (Fig.~\ref{Fig4}c).
\begin{figure*}
  \centering
    \includegraphics[trim= 0 0 0 0, clip,width=0.999\textwidth]{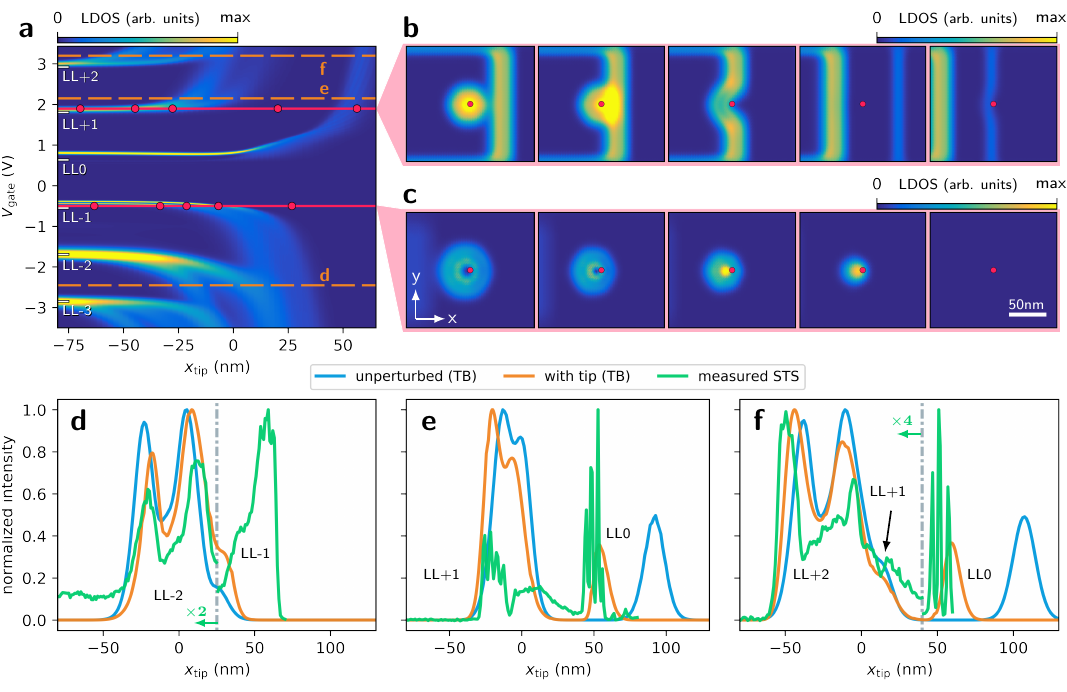}	
    \caption{
    {\bf Quantitative comparison between scanning tunneling spectroscopy data and TB simulations:}
    (a) Same data set as in Fig.~\ref{Fig2}d with different red lines and dots  according to b--c and additional dashed orange lines indicating $V_{\rm gate}$ of the line profiles in d--f.
    (b), (c) LDOS$(x,y)$ at the $V_{\rm gate}$ as marked in a and $\mathbf{x}_{\rm tip}$ marked by the red dot in each panel as well as in a.
{(d)--(f)} Profile lines along the dashed lines in a (orange), across the calculated LDOS at the same $V_{\rm gate}$, but without the TIQD (blue, Appendix Section~\ref{sec:PoissontoTB})
and across the experimental data of Fig.~\ref{Fig1}g at the same $V_{\rm gate}$ (green). The experimental profiles have been horizontally shifted by $+5$\,nm (d), $+20$\,nm (e), $+10$\, nm (f) to ease the comparison. Additional intensity adjustments as marked are used to compensate for the strong charging lines that are not included in the simulation. The peak fine structure is a fingerprint of the charging lines (compare Fig.~\ref{Fig1}g)
}
    \label{Fig3}
\end{figure*}
\subsection{Tight Binding Simulations Including the Tip Induced Quantum Dot}
\label{sec:TB_TIQD2}
In order to explain the observed branching of various LL at the interface, we perform TB simulations employing a realistic potential. The potential is deduced from Poisson simulations with parameters extracted from the experimental charging lines. We consider the work function mismatch between graphene and the tip as well as between graphene and the graphite gate  and (assuming an approximately spherical tip) the radius of the tip apex $r_{\rm tip}$ (Appendix Section~\ref{sec:Poissonsim}). The work function mismatches are quantified as voltages $ \Delta V_{\rm gate}$ and $\Delta V_{\rm sample}$ required for charge neutrality of the graphene and flat band conditions below the tip, respectively. These parameters can be deduced from the first crossing points of charging lines originating from adjacent LL$n$ (Fig.~\ref{Fig2}a). Such a crossing implies a potential depth of the TIQD identical to the energy difference between the two LL$n$ (Fig.~\ref{FigS1}d, appendix). Using two such crossing points at two pairs of ($V_{\rm sample}$, $V_{\rm gate}$), we straightforwardly determine $\Delta V_{\rm gate}=-200\pm 50$\,mV and $\Delta V_{\rm sample}=-180\pm 50$\,mV by comparison with Poisson simulations (Appendix Section~\ref{sec:ParametersPoisson}). The parameter $r_{\rm tip}=25$\,nm is deduced from the average distance of charging lines again by comparison with the Poisson simulations (Appendix Section~\ref{sec:ParametersPoisson}).

Two resulting TIQD potentials (far away from the lateral interface) are shown in Fig.~\ref{Fig2}b.
The complete lateral potential $\Phi_{\rm tot} (x)$ including TIQD and interface potential  results from a Poisson simulation using the geometry of Fig.~\ref{Fig1}c as well as the known $V_{\rm sample}$, $V_{\rm gate}$, $x_{\rm tip}$, $\Delta V_{\rm gate}$, $\Delta V_{\rm sample}$, and $r_{\rm tip}$ (Appendix Section~\ref{sec:PoissonlatInt}). The potential features multiple steps  across the interface due to alternating compressible and incompressible stripes (Fig.~\ref{Fig2}c, lower and upper frame).\cite{Chklovskii1992,Lier1994}

These potentials, adequately transformed into 2D potentials (Appendix Section~\ref{sec:PoissontoTB}), are the input for the TB simulations.
We construct the LDOS from the resulting single particle states: we sum all states around the energy selected by $V_{\rm sample}$ in an energy window of $\sim 3$\,meV in order to capture the temperature broadening in the experiment.\cite{Morgenstern2000b} Spatially, we average across a circular region (radius $\approx 1.5$ nm) around the 2D tip center position $\mathbf{x}_{\mathrm{tip}}$ to account for  a possible mismatch between the tunneling position and the capacitive center of the tip. \cite{Morgenstern2000} Figure~\ref{Fig2}d shows the resulting LDOS for a direct comparison to Fig.~\ref{Fig1}g. Crucially, the branching features of the various LL$n$ states are correctly reproduced, while LL0 does not exhibit any branching.
The favorable agreement calls for a detailed study of the complete LDOS$(x,y)$ map at various $\mathbf{x}_{\rm tip}$ as naturally provided by the TB simulations. The calculations reveal that the branching is a consequence of the internal structure of the edge state wave functions at the interface (Fig.~\ref{Fig2}e--g). For example, the edge state belonging to LL-2 (Fig.~\ref{Fig2}g) exhibits two antinodes that are probed by the tip as two arms of a branching of the LL-2 state (Fig.~\ref{Fig2}d). By contrast, the edge state of LL0 with a single antinode (Fig.~\ref{Fig2}f) does not show branching in the probed LDOS of Fig.~\ref{Fig2}d.
A local displacement of the edge state by the TIQD potential (apparent in Fig.~\ref{Fig2}e--g) only shifts the lateral position of the edge state center with minor influence on its internal structure (see below).

Figure~\ref{Fig3}b--c reveals instead that the intense horizontal LDOS lines  observed to the far left of the lateral interface (Fig.~\ref{Fig3}a) are caused by states of the TIQD. These states are shifted in energy across the interface 
and, hence, can disappear from the probed energy window. Thus, only the weaker LDOS features across the interface contain the desired edge state information.
\begin{figure*}
    \centering
    \includegraphics[width=0.999\textwidth]{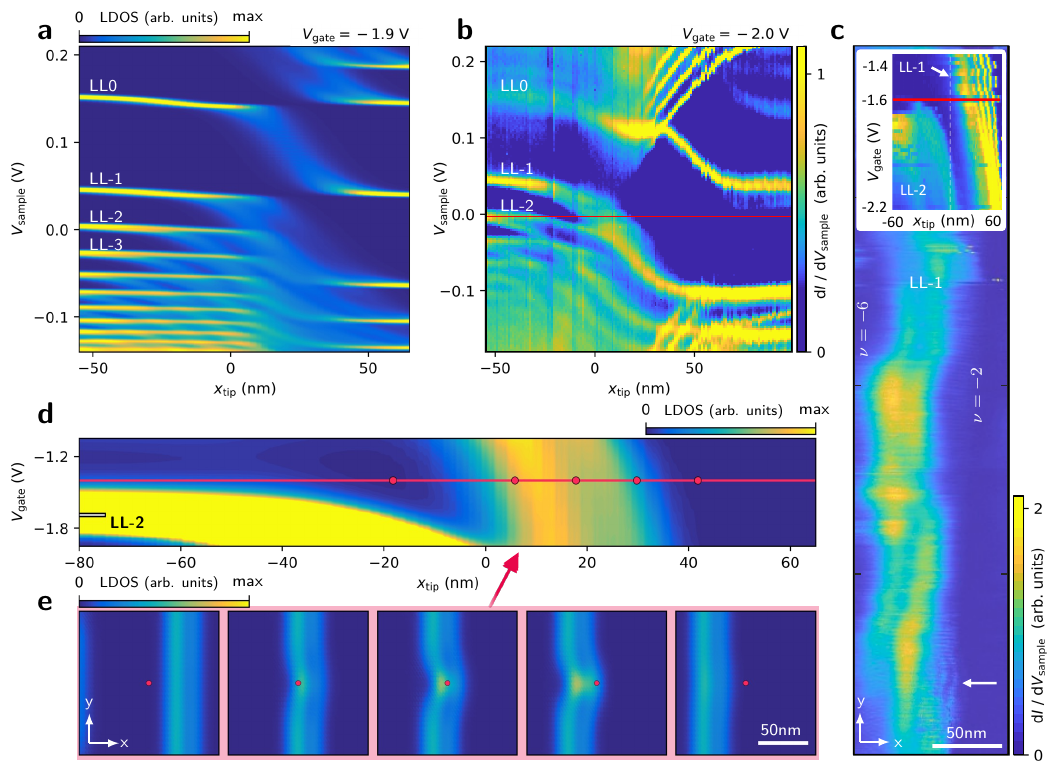}
    \caption{
        {\bf Mapping of edge state along the interface:}
        (a) Simulated LDOS($x_{\rm tip}$,\,$V_{\rm sample}$,) across the lateral interface, 
        while including the TIQD, 
        $V_{\rm{gate}}=-1.9$\,V.
        (b) Measured $dI/dV_{\rm sample}( x_{\rm tip},\,V_{\rm sample})$, $V_{\rm gate}=-2.0$\,V, $I_{\rm stab}=200$\,pA, $V_{\rm stab}=-250$\,mV.
        (c) $dI/dV_{\rm sample}(\mathbf{x}_{\rm tip})$ along the lateral interface ($y$ direction) featuring the LL-1 edge state at $E_{\rm F}$, $V_{\rm gate}=-1.6\,\rm V$, $V_{\rm sample}=0\,\rm V$, $I_{\rm stab}=1$\,nA, $V_{\rm stab}=-250$\,mV. Filling factors $\nu$ are marked on both sides of the interface. Inset shows $dI/dV_{\rm sample}( x_{\rm tip},\,V_{\rm gate})$ recorded across the interface at the position marked by a white arrow in the main image. The red line marks $V_{\rm gate}$ of the main image.
        (d) Zoom into Fig.~\ref{Fig3}a (simulated LDOS, $V_{\rm sample}=0$\,V) at larger contrast to visualize the internal structure of the LL-1 edge state.
        (e) Simulated LDOS$(x,y)$ across the interface for various $\mathbf{x}_{\rm tip}$ marked by red dots (also in d), $V_{\rm{gate}} = -1.4$\,V (red line in d).}
    \label{Fig4}
\end{figure*}

\subsection{Direct Comparison of Experimental and Simulated Data}
\label{sec:compare}

To elucidate the remaining influence of the TIQD, we now directly compare the calculated cross-section of the LDOS related to the edge states (blue lines in Fig.~\ref{Fig3}) to the measured $dI/dV_{\rm sample}(x_{\rm tip})$ (green lines) and to the simulated $dI/dV_{\rm sample}(x_{\rm tip})$ including the TIQD (orange lines). Favorably, the twofold an\-ti\-no\-dal structure of LL-2 (Fig.~\ref{Fig3}d) and LL+2 (Fig.~\ref{Fig3}f) appears very similarly in all three curves, i.e. peak distances and relative intensities are alike. This good agreement for LL-2 can be traced back to the fact that the TIQD is absent at the interface region (Fig.~\ref{Fig2}c, lower frame,$-20\,$nm$< x_{\rm tip}<20\,$nm). Analyzing the distance of antinodes $\Delta x$ in more detail reveals $\Delta x =31\pm 1$\,nm in the experiment largely independent of $V_{\rm gate}$. In the TB calculations with TIQD, we find $\Delta x =23\pm 2$\,nm  slightly decreasing with increasing $V_{\rm gate}$. A 1D TB calculation representing the overlapped unperturbed LL wave functions of the two sublattices finds $\Delta x =25$\,nm (Appendix Section~\ref{sec:Comparison}), i.e. the experimental distance of antinodes is larger by $\sim 25$\,\%. Slightly larger distances in the experiment are also observed for LL+1 (Fig.~\ref{Fig3}e), with values largely independent from $V_{\rm gate}$ in experiment ($\Delta x =25\pm 2$\,nm) and simulations ($\Delta x =15\pm 1$\,nm) (Appendix Section~\ref{sec:Comparison}), and for LL+2 (Fig.~\ref{Fig3}f). The larger distances in the experiment could be due to the neglected electron-electron repulsion in the TB calculation. In a perturbation theory approach, Coulomb repulsion would mix the states at $E_{\rm F}$ with higher LLs, compensating the energy cost for partial occupation of higher LLs by the gain in Coulomb energy due to the increased lateral extension of the states at $E_{\rm F}$. Finally, we note the slight variation in peak heights of the LL edge states in Fig.~\ref{Fig3}d-f. For example, the first  of the two peaks of the LL-2 feature in Fig.~\ref{Fig3}d is slightly lower in intensity than the second in both experiment and theory. This variation is caused by, both, the presence of the TIQD (compare relative peak heights of the orange and blue lines), and by the finite slope of the potential at the interface (Appendix Section~\ref{sec:Comparison}).

Discrepancies in relative intensities of antinodal peaks get, however, significant, if charging lines interfere (LL0 in Fig.~\ref{Fig3}e--f).  The peak distances in Fig.~\ref{Fig3}e still match reasonably between green and orange lines, but not the relative peak intensities. More severely, the distances between the edge state peaks of LL+1 (LL+2) and LL0 (Fig.~\ref{Fig3}e (f)) are considerably reduced by the presence of the TIQD (blue vs. orange lines). This relates to a shift of the most right incompressible stripe towards the left by the superposed TIQD potential (Fig.~\ref{Fig2}c, upper frame). Nevertheless, the simulated shift of LL0 by the TIQD (orange) is in quantitative agreement with the experiment (green). The fact that the LL-1 feature in Fig.~\ref{Fig3}d strongly deviates from the simulated peak in terms of position and intensity is likely related to the interfering charging lines (Fig.~\ref{Fig1}g) of a relatively shallow TIQD (Fig.~\ref{Fig2}c, lower frame, $ x_{\rm tip}>40\,$nm). In such a shallow TIQD, individual charging events can strongly change the TIQD potential, an effect not captured by the simulations. Thus, imaging of the edge states works best if no charging lines are observed in the corresponding parameter regime and the TIQD is absent in the region of the lateral interface.

\subsection{Mapping the Edge State}
\label{sec:edgemap}
To corroborate the generally good agreement between measured $dI/dV_{\rm sample}$ and simulated LDOS,
we compare their dependence on $V_{\rm sample}$ and $x_{\rm tip}$ in Fig.~\ref{Fig4}a--b (see also Fig.~\ref{FigS7}). Again, one observes semi-quantitative agreement including the branching features of LL-1 to LL-3. At a  slightly smaller $V_{\rm gate}$, only LL-1 crosses $E_{\rm F}$ at the interface (red line in inset of Fig. \ref{Fig4}c). At this $V_{\rm gate}$, we map $dI/dV_{\rm sample}(\mathbf{x}_{\rm tip})$ two-dimensionally at $E_{\rm F}$ (Fig.~\ref{Fig4}c). A bright line about 40\,nm in width with some internal structure meanders along the lateral interface. Mostly, the bright line is a double line structure as expected for LL-1 (compare TB simulation of Fig.~\ref{Fig3}e showing LL+1) Width and internal structure of this stripe are rather similar to the simulated LDOS($x_{\rm tip}$) of the LL-1 edge state (Fig.~\ref{Fig4}d, along red line). An analysis of the correspondingly mapped LDOS (Fig.~\ref{Fig4}e) reveals that the observed double line is due to the intrinsic double line of the LL-1 edge state, itself caused by the antinodal structure of the LL-1 wave function that is  barely perturbed  by the shallow TIQD. In some areas, as marked by the white arrow in Fig.~\ref{Fig3}c, we observe additional charging lines (see also inset) due to the local potential that changes the occupation of the TIQD, but these areas are small at the chosen $V_{\rm gate}$.

Hence, an imaging of an edge state with resolution well below the magnetic length $l_{\rm B}=10$\,nm and only minor perturbations by the TIQD is achieved for the first time.

\section{Conclusions}
\label{sec:Conclusion}
We conclude that  quantum Hall edge states can be mapped without significant perturbations if one selects favorable parameter regimes. One attractive option to identify such regions is a direct comparison of $dI/dV_{\rm sample}$ across a gated lateral interface with TB simulations accounting for the TIQD. Crucially, reliable parameters for simulating the TIQD can be straightforwardly deduced from the measured charging lines in $dI/dV_{\rm sample}(V_{\rm gate},\,V_{\rm sample})$. Current limitations of the method include neglecting confinement effects on the shape of the TIQD, which would require more time-consuming Poisson-Schr\"odinger simulations and, probably more severe, the assumption of a circularly symmetric TIQD. Trial and error-type control on the TIQD shape, however, can generally be achieved by mapping the capacitive charging of a point defect.\cite{Morgenstern2000,Teichmann2008} Even with these limitations, the antinodal structure of the edge states could be mapped in a largely quantitative fashion, even revealing the influence of the potential gradient at the interface on the relative peak heights. An additional gate that can also tune the filling factor on the other side of the interface might eventually give access to multiple nearly unperturbed edge states including some that separate symmetry broken \cite{Wang2022} or fractional QH phases.

During the final preparation of this manuscript, we became aware of measurements attempting to probe quantum Hall edge states at the physical edge of graphene on hBN/SiO$_2$/Si. They did not find signatures of edge states, again likely due to a too strong edge potential.\cite{Coissard2022}

\section*{Author Contributions}
T.J. provided the idea of the experiment and performed the experimental measurements as well as the Poisson simulations, supervised by M.M. and S.S.. A.W. and M.H prepared the sample supervised by R.G..
C.S. provided the tight-binding simulations supervised by F.L..
M.M. conceived and supervised the project.
All authors discussed the results and co-wrote the manuscript.

\section*{Competing Interests}
The authors declare no competing interests.

\section*{Acknowledgements}
We gratefully acknowledge helpful discussions with T. Fabian, V. Falko and M. Goerbig. This project has received funding from the European Union's Horizon 2020 research and innovation programme under grant agreement number 881603 (Graphene Flagship, Core 3) and by the German Science foundation via Mo 858/16-1 as well as Mo 858/15-1. We further acknowledge support from the FWF DACH project I3827-N36 and COST action CA18234. Christoph Schattauer acknowledges support as a recipient of a DOC fellowship of the Austrian Academy of Sciences. Numerical calculations were in part performed on the Vienna Scientific Cluster VSC4. Roman Gorbachev acknowledges support from Royal Society, ERC Consolidator grant QTWIST (101001515) and EPSRC grant number EP/V007033/1.

\section{Appendix}
\label{sec:appendix}
\subsection{Sample Preparation, Locating the Lateral Interface by STM and Measuring the STS Signal}
\label{sec:prep_sample}
\begin{figure}
    \centering
    \includegraphics[scale=0.5]{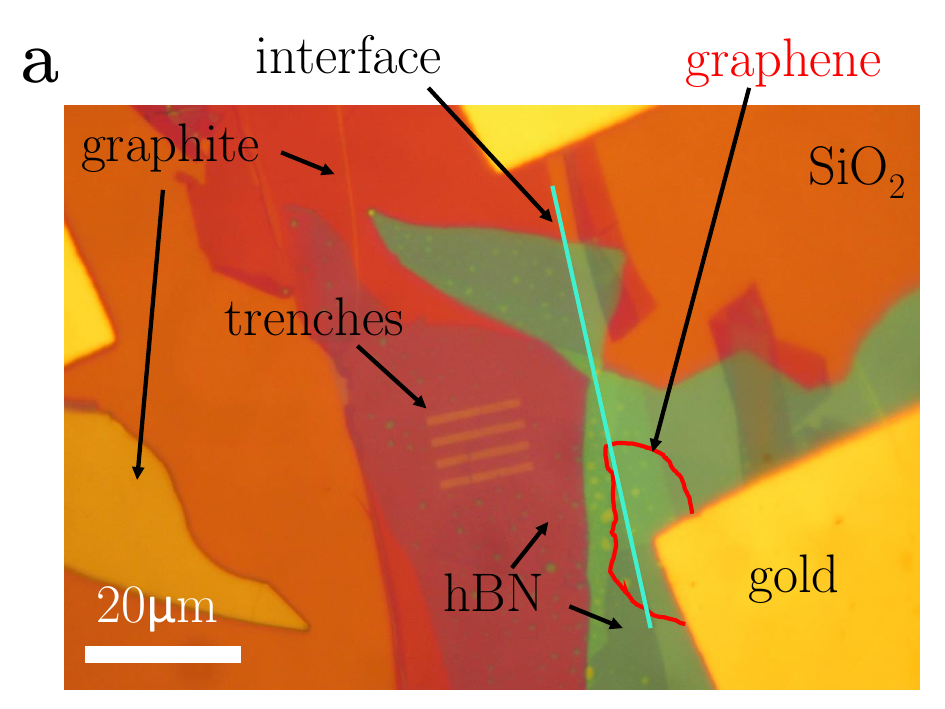}
    \includegraphics[scale=0.5]{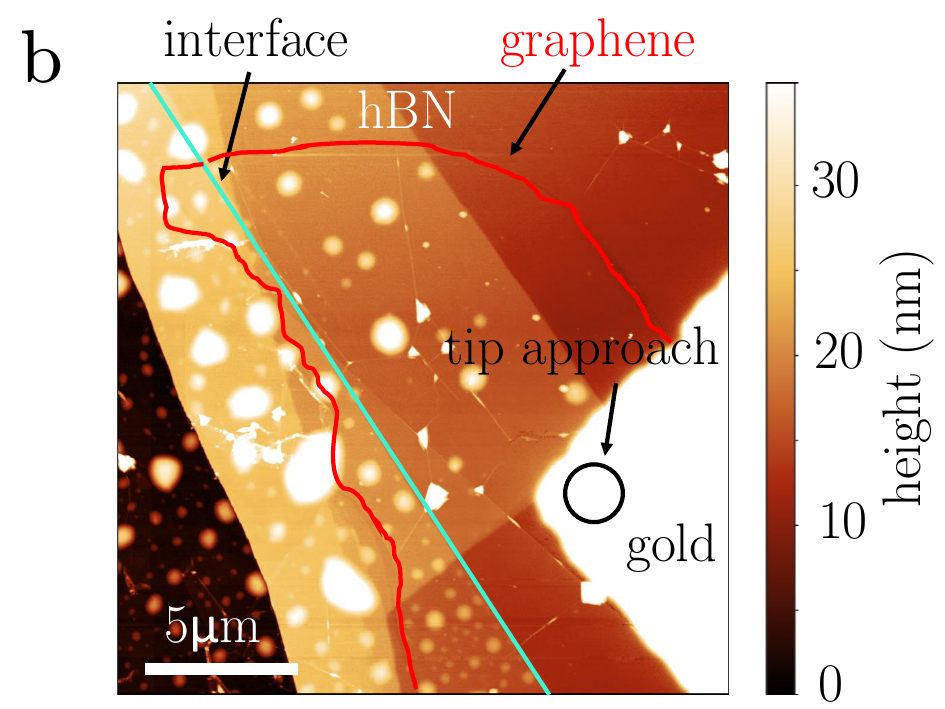}
     \includegraphics[scale=0.5]{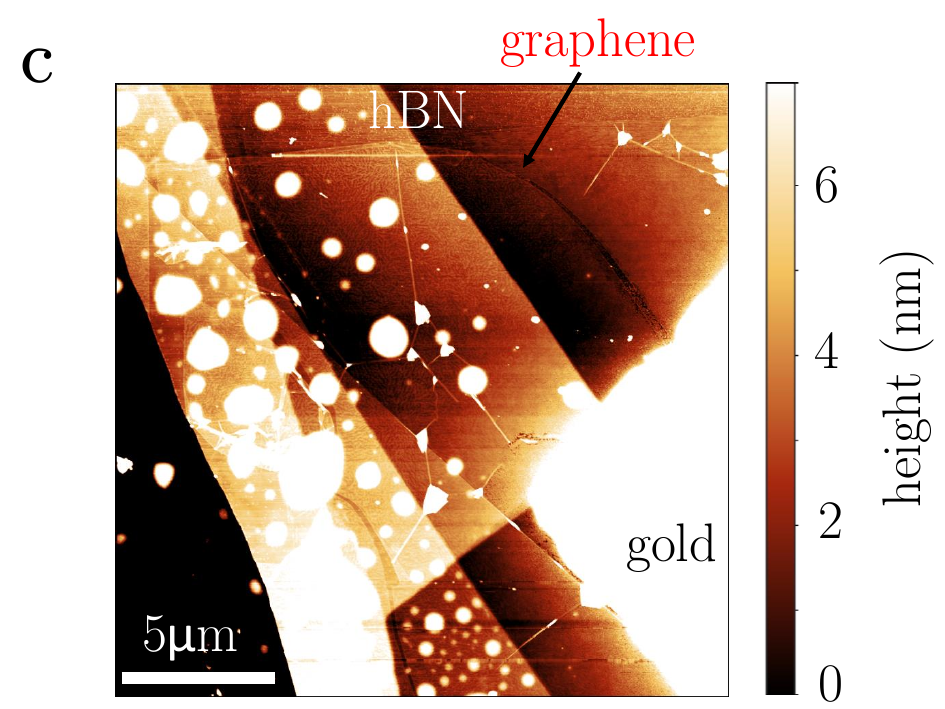}
    \caption{
     (a) Optical image of the finalized sample with different areas marked. The graphene area  is encircled (red line) as determined by atomic force microscopy (AFM). The graphene has been ruptured during the transfer, such that the trenches within the graphite are not used. The studied lateral interface is marked (blue line) separating the graphite gate area on the left and an area without graphite gate on the right.  (b) AFM image acquired in tapping mode at ambient conditions, Si-cantilever, $f_{\rm res}=325$\,kHz. The rim of the contacted graphene flake is marked by a red line. The circle indicates the intended landing position of the STM tip. (c) Same as b with different plane fit and different contrast such that the rim of the graphene flake gets partly visible. The flake is ruptured such that only the small groove between the contacted graphene area and the surrounding graphene is visible.
    }
    \label{Fig7}
\end{figure}

The sample is prepared by firstly exfoliating a graphite flake onto a Si/SiO$_2$ chip.  Afterwards, two hBN flakes and a graphene flake are transferred onto the graphite  by dry stacking such that the graphene only partially covers the graphite.\cite{Kretenin2014} Finally, the graphene and the graphite are electrically contacted. For this purpose, a shadow mask is fixed on the chip through which 60\,nm high gold contacts are evaporated by a thermal gold evaporator. An optical image of the sample is shown in Fig.~\ref{Fig7}a and atomic force microscopy (AFM) images are provided in Fig.~\ref{Fig7}b--c. The chip with the sample is then glued onto a STM sample holder.  The graphene flake and the Si back gate are connected to the sample holder with silver paint. One additional contact connects the graphite flake independently via a gold wire. After loading the sample into the STM of the ultrahigh vacuum chamber, the tip is positioned onto the corner of the gold contact next to the graphene flake (Fig.~\ref{Fig7}b) using an optical long-distance microscope for monitoring. Then, the tip is approached until tunneling current is achieved. The tip is afterwards moved laterally towards the graphene flake while continuously recording the topography. It is easy to recognize the graphene, since the gold is significantly more rough. With the tip on the graphene, the lateral interface is eventually identified by a 3\,nm high step  resulting from the underlying graphite (Fig.~\ref{Fig1}d, main text).

For STM and scanning tunneling spectroscopy (STS),  we used an etched W wire that is prepared on W(110) by voltage pulses until a reliable $I(z)$ curve and stable $dI/dV_{\rm sample}(V_{\rm sample})$ were obtained. After maneuvering the tip to the graphene, additional mild voltage pulses are applied on the Au contact pads next to the graphene to get rid of possible dirt that is picked up during the path towards the graphene.
We did not use other tip materials to tune the tip work function, since variations of the work function for the same material  due to different facets at the tip apex can amount to up to 250\,meV already.\cite{Dombrowski1999}
For STS, the voltage $V_{\rm sample}$ is applied to the graphene and the tunneling current $I$ is recorded at the tip. Lock-in technique probes $dI/dV_{\rm sample}(V_{\rm sample})$ after opening the feedback loop at voltage $V_{\rm stab}$ and current $I_{\rm stab}$. The modulation frequency is $f={386.2}$\,Hz for all images except Figs.~\ref{Fig1}f, \ref{Fig4}b, main text, and Figs. \ref{FigS7}b, \ref{Fig_comp_split}, where we used $f=1386$\,Hz. The modulation amplitude is $V_{\rm mod}^{\rm rms}=1$\,mV except for Figs.~\ref{Fig1}f, \ref{Fig4}b, main text, and Figs. \ref{FigS7}b, \ref{Fig_comp_split},  where it is $V_{\rm mod}^{\rm rms}=5$\,mV.

\subsection{Poisson Simulation}
\label{sec:Poissonsim}
To perform tight binding (TB) calculations for comparison with  STS, we need the 2D potential profile on the graphene around the interface.
It consists of the potential step induced by the graphite gate
and the  tip induced quantum dot (TIQD)
caused by the potential difference between tip and sample.
The potentials are due to the applied voltages $V_{\rm gate}$ between graphite gate and graphene as well as $V_{\rm sample}$ between the tip and graphene (Fig.~\ref{Fig1}c, main text) and the corresponding work function mismatches. Moreover, the geometry of these two metallic electrodes and the density of states (DOS) of the graphene are relevant.

For estimating the resulting potential,
we employ numerical Poisson calculations. The home-made Poisson solver uses either 2D Cartesian coordinates or coordinates for a 3D cylindrical symmetry. The calculations disregard confinement effects, i.e. we do not use a Poisson-Schr\"odinger solver. Instead, we treat the DOS as a property that is rigidly shifted by the local potential. The Landau level structure of the DOS as well as the temperature via the Fermi-Dirac distribution are taken into account.\cite{Freitag2016}

\begin{figure}
    \centering
    \includegraphics[scale=0.59]{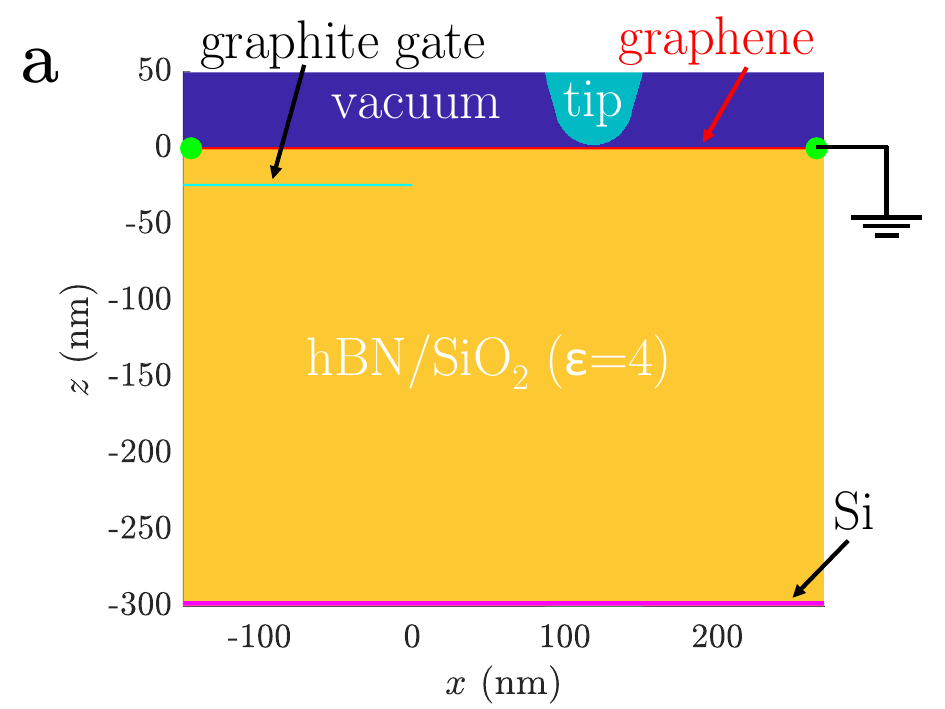}
    \includegraphics[scale=0.59]{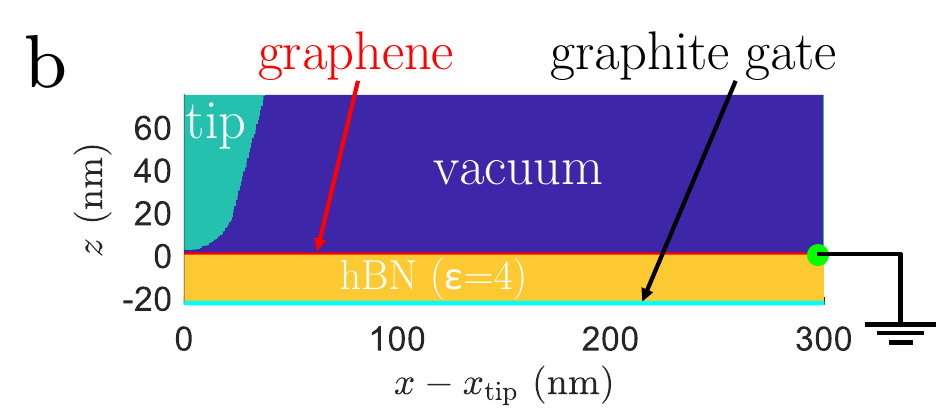}
    \caption{
     (a) Two-dimensional Cartesian geometry as used for the Poisson simulations that include the lateral interface.  Different regions are marked. The potentials $\Phi_{\rm tip}$ and $\Phi_{\rm gate}$ are applied to the tip and the graphite gate, respectively. The thickness of the graphite is omitted for the sake of simplicity.
    (b) Cylindrical geometry as used for Poisson simulations of the TIQD without lateral interface using a graphite gate that covers the complete lateral area. The geometry is taken as rotationally symmetric.
}
    \label{FigS4}
\end{figure}

Fig.~\ref{FigS4} shows the chosen geometries for the Poisson simulations. They are largely identical to Fig.~\ref{Fig1}c, main text, except that the graphite is  two-dimensional without extension in $z$-direction.
The thickness of the hBN $d_{\rm hBN}=23.5$\,nm is deduced from atomic force microscopy (AFM) images. Both, hBN and SiO$_2$ are described by their dielectric bulk constant $\epsilon\simeq 4$. We choose a reasonable value  $d_{\rm tip}= 0.6$\,nm for the distance between graphene and the tip apex since it barely influences the results.\cite{Freitag2016,Freitag2018,Morgenstern2000b}
The tip is assumed to be metallic with a shape consisting of a half sphere with radius $r_{\rm tip}$ located at the lower end of a cone with opening angle $30^\circ$.

The graphite is set to a potential $\Phi_{\rm gate}=e\cdot (V_{\rm gate}-\Delta V_{\rm gate})$, where $e=1.6\cdot10^{-19}$\,C. $V_{\rm gate}$ is the applied gate voltage and $\Delta V_{\rm gate}$ is the required gate voltage to achieve charge neutrality in the graphene.
In the Poisson simulations, the  sample is grounded and
the tip is set to a variable potential $\Phi_{\rm tip}$. This is different from the experiment, where the tip is grounded. The reason is that the sample grounding at the graphene edge enables a more straightforward implementation of the Poisson solver.\cite{Freitag2016}
The tip potential, hence, reads $\Phi_{\rm tip}=e\cdot (-V_{\rm sample}+\Delta V_{\rm sample}$) with applied graphene voltage $V_{\rm sample}$ and
$\Delta V_{\rm sample}$ being the voltage required to achieve flat band conditions below the tip for $\Phi_{\rm gate}=0$\,eV.

The used Landau level (LL) DOS employs a Fermi velocity $v_{\rm F}=1\cdot 10^6$\,m/s leading to LL energies:\cite{CastroNeto2009}
\begin{equation}
E_{\mathrm{LL}n} = v_{\rm F} \cdot \mathrm{sgn}(n) 
\sqrt{2 \hbar e|B| |n|}, \hspace{0.5cm}   \hspace{0.5cm} n \in \mathbb{Z},
\label{relLandau}
\end{equation}
with $B$ the magnetic field perpendicular to the graphene and $n$ the Landau level index.
A Gaussian broadening of the Landau levels with FWHM of $7$\,meV is additionally applied. 

\begin{figure}
    \centering
    \includegraphics[scale=0.59]{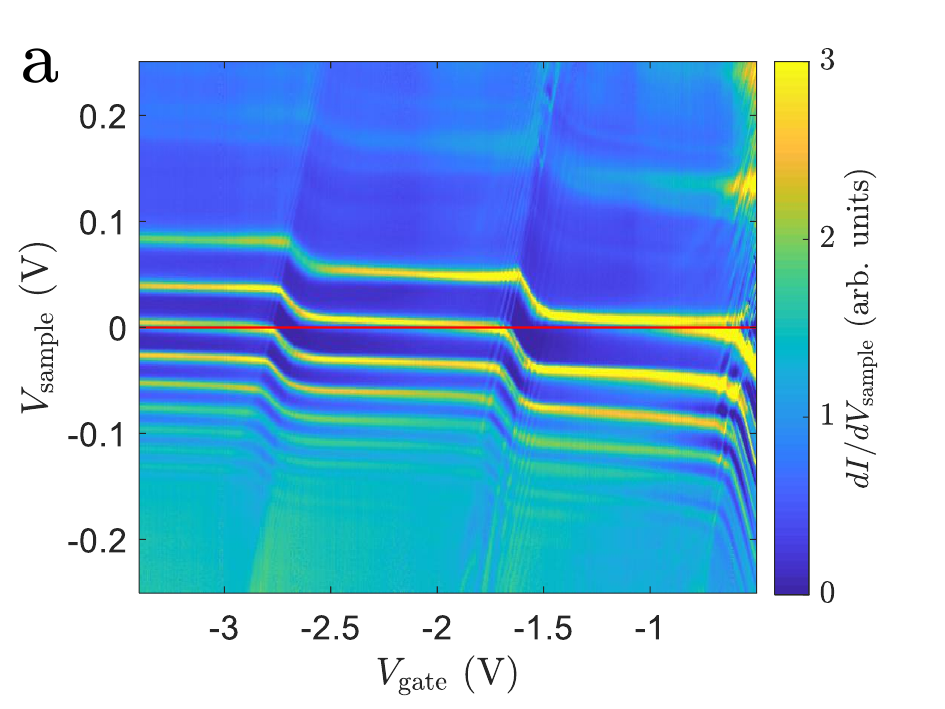} \includegraphics[scale=0.59]{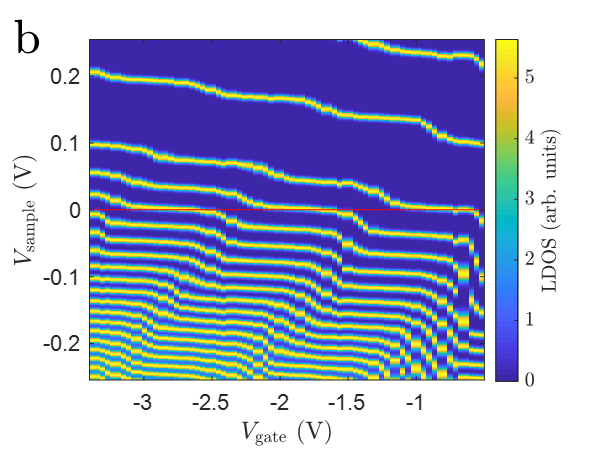}
    \includegraphics[scale=0.59]{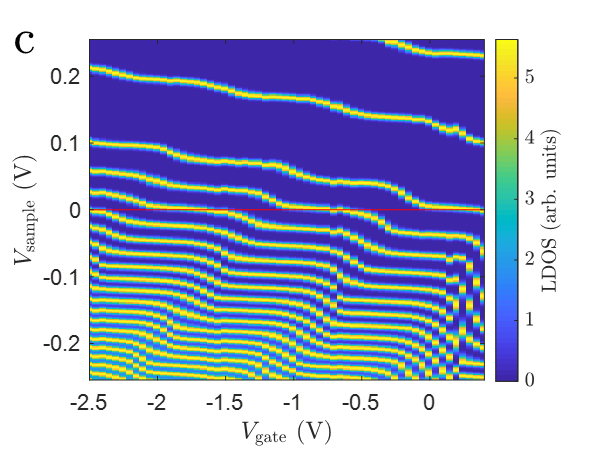}
    \vspace{-0.8cm}
    \caption{(a) $dI/dV_{\rm sample}$ ($V_{\rm gate}$, $V_{\rm sample}$) recorded at a position $x_{\rm tip} \ll 0$\,nm, $I_{\rm stab}=1$\,nA, $V_{\rm stab}=-250$\,mV. (b)
    LDOS($V_{\rm gate}$, $V_{\rm sample}$) resulting from the Poisson simulations with optimized parameters, $\Delta V_{\rm gate}=-200$\,mV, $\Delta V_{\rm sample}=-180$\,mV, $r_{\rm tip} = 25$\,nm.
    (c)
    LDOS($V_{\rm gate}$, $V_{\rm sample}$) resulting from the Poisson simulations with less favorable  parameters, $\Delta V_{\rm gate}=+650$\,mV, $\Delta V_{\rm sample}=-230$\,mV, $r_{\rm tip} = 25$\,nm. Note the shifted $V_{\rm gate}$ axis in c.
    }
    \label{FigS3}
\end{figure}

\begin{figure}
    \centering
    \includegraphics[scale=0.59]{sf_STS_compare_exp.pdf} \includegraphics[scale=0.59]{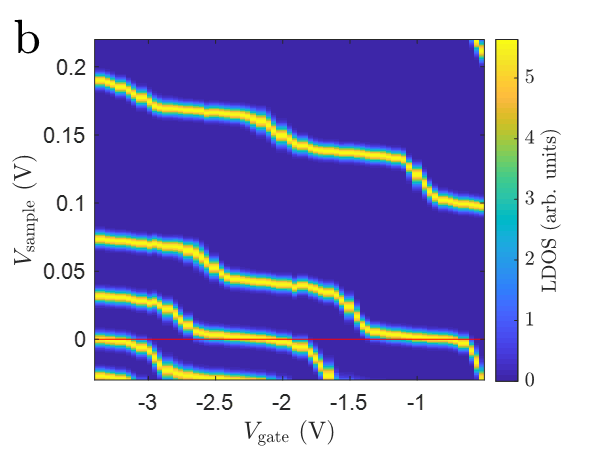}
    \vspace{-0.8cm}
    \caption{(a) $dI/dV_{\rm sample}$ ($V_{\rm gate}$, $V_{\rm sample}$) recorded at a position $x_{\rm tip} \ll 0$\,nm, $I_{\rm stab}=1$\,nA, $V_{\rm stab}=-250$\,mV. (b)
    LDOS($V_{\rm gate}$, $V_{\rm sample}$) resulting from the Poisson simulations with optimized parameters, $\Delta V_{\rm gate}=-200$\,mV, $\Delta V_{\rm sample}=-180$\,mV, $r_{\rm tip} = 25$\,nm at a modified $\epsilon = 3$ (Fig.~\ref{FigS3}b: $\epsilon = 4$).
    }
    \label{Fig1_epsilon}
\end{figure}

We eventually plot a simulated LDOS($V_{\rm gate}$, $V_{\rm sample}$) derived from the LDOS of graphene directly below the tip center at the energy with respect to the Fermi level of the sample that matches $V_{\rm sample}$ (Fig.~\ref{FigS3}b, c). This enables a comparison with the measured $dI/dV_{\rm sample}$ ($V_{\rm gate}$, $V_{\rm sample}$) (Fig.~\ref{FigS3}a)  and, hence, an optimization of parameters (see below).

The agreement between Fig.~\ref{FigS3}a and c is reasonable, i.e the same LLs are crossing $E_{\rm F}$ in the experimental range at similar $V_{\rm gate}$. However, details are different, e.g. the lengths of the plateaus are shorter in the simulation. This can be improved by adapting $\epsilon$ as an additional fit parameter (Fig.~\ref{Fig1_epsilon}) taking into account that the dipolar screening of hBN and SiO$_2$ is modified at its surfaces. We omit such additional fit parameter in the following to keep the number of fit parameters low and, thus, the reasoning more transparent.

\subsection{Determining the Parameters for the Poisson Simulation}
\label{sec:ParametersPoisson}
As decisive parameters for the Poisson simulations, we need to determine $\Delta V_{\rm sample}$, $\Delta V_{\rm gate}$ and $r_{\rm tip}$.
For this purpose, we use the observed charging lines in $dI/dV_{\rm sample}(V_{\rm gate},V_{\rm sample})$ at
$x_{\rm tip} \ll 0$\,nm, i.e  far away from the lateral interface  (Fig.~\ref{Fig1}a-b, main text). The charging lines are directly related to the TIQD potential.
The tip radius $r_{\rm tip}$ correlates with the lateral size of the TIQD, i.e. with the distance of charging lines, whereas $\Delta V_{\rm sample}$ and $\Delta V_{\rm gate}$ affect the potential depth of the TIQD, i.e. the onset of charging lines for each LL$n$.
\begin{figure}
    \centering
    \includegraphics[scale=0.35]{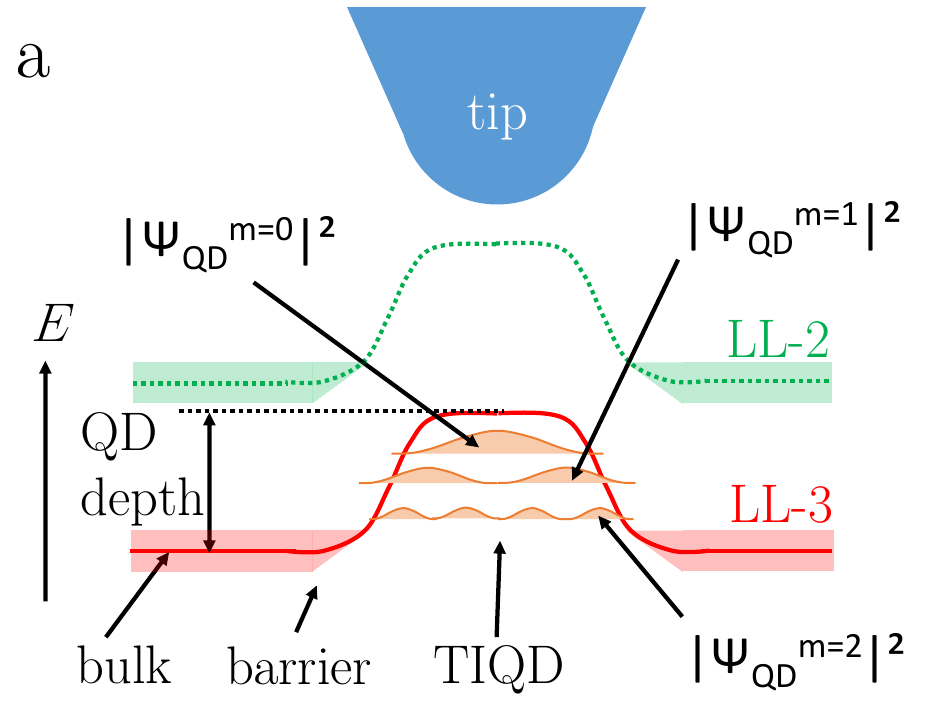}
    \includegraphics[scale=0.35]{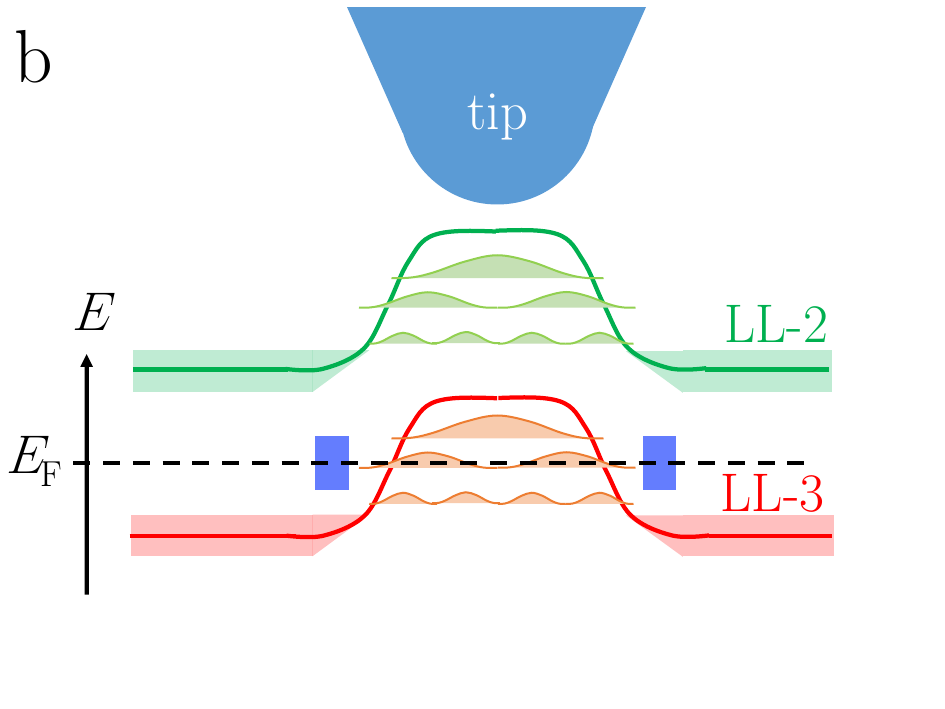}
    \includegraphics[scale=0.35]{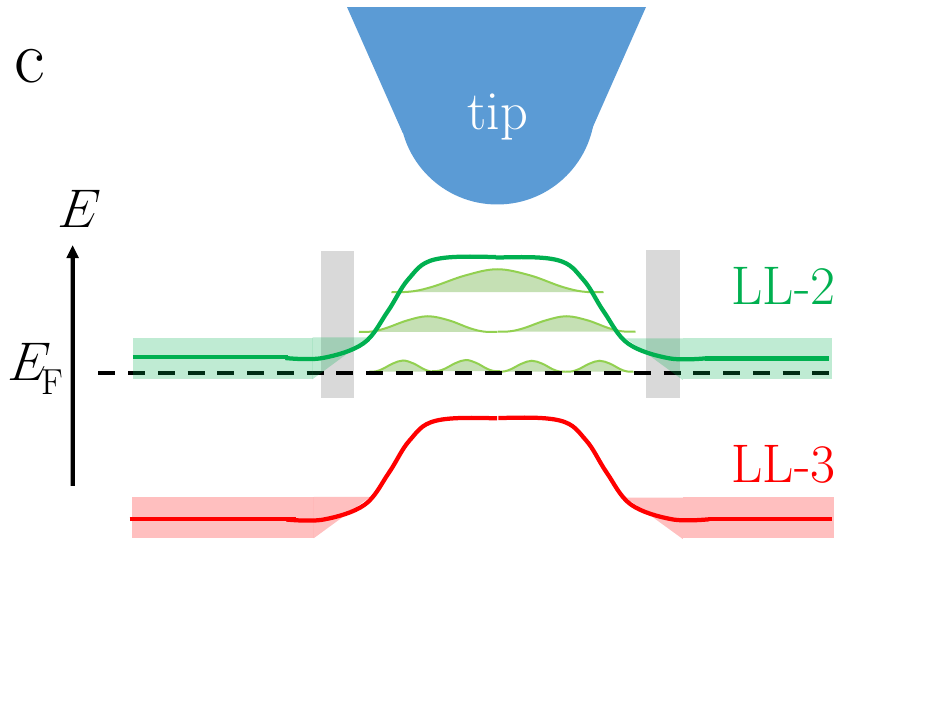}
    \includegraphics[scale=0.35]{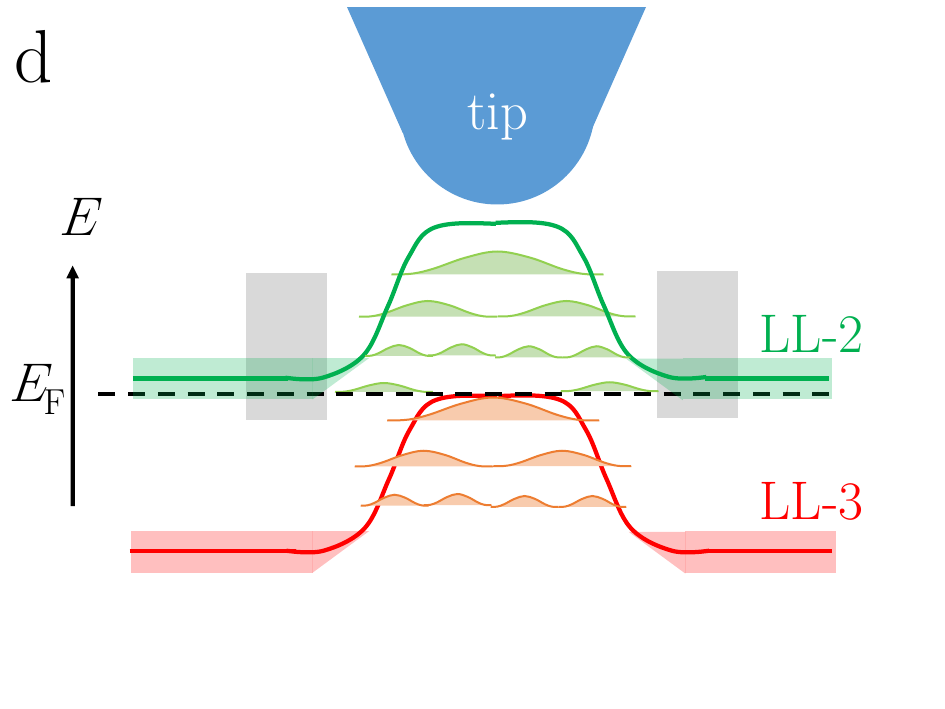}
    \vspace{-1cm}
\caption{
     (a) Potential energy course of two adjacent LL$n$ due to the TIQD potential. The probability density of the confined states of LL-3 are added at their confinement energy and labeled with its azimuthal quantum number $m$. The ($m=0$)-state provides the largest probability density in the center of the TIQD. (b) Same as a with added $E_{\rm F}$ line (dashed) at the ($m=1$)-state of LL-3. The blue shaded areas mark the onset of the insulating surrounding of the TIQD.   (c) Similar sketch with $E_{\rm F}$ at the charging position of the last state from LL-2. The indicated grey tunnel barrier onsets originate from localized states in the bulk of LL-2 surrounding the TIQD. (d) Sketch with two states at $E_{\rm F}$ enabling simultaneous charging of the first hole of LL-3 and the last hole of LL-2. Then, the depth of the TIQD roughly equals the energy distance between LL-2 and LL-3.
}
    \label{FigS1}
\end{figure}

 \begin{figure}
    \centering
    \includegraphics[scale=0.5]{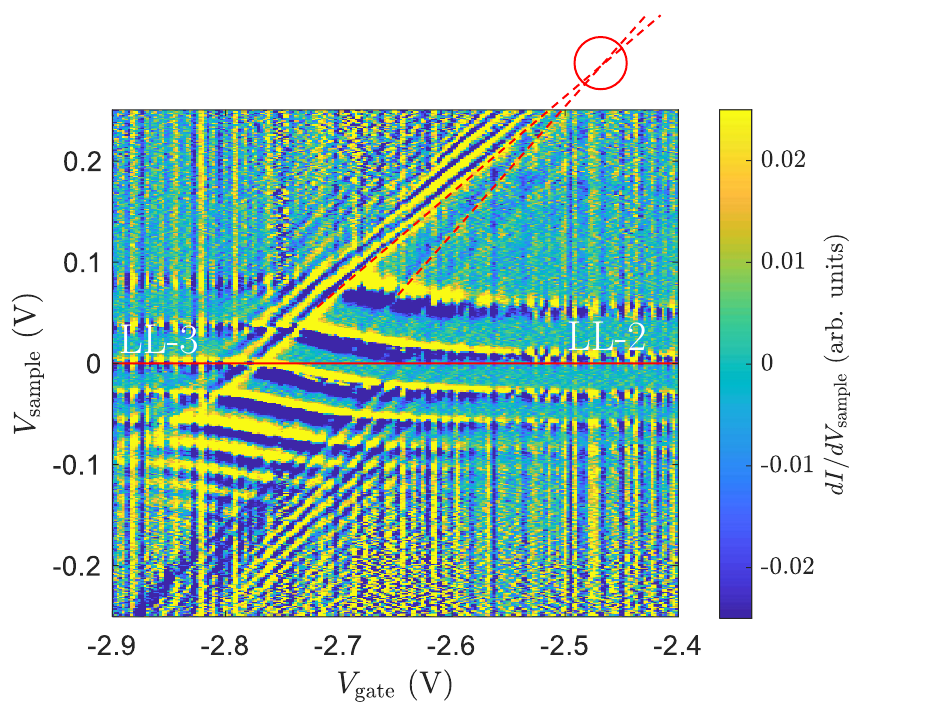}
    \caption{
$d^2I/dV_{\rm sample}dV_{\rm gate}(V_{\rm gate},\,V_{\rm sample})$ for the transition between LL-3  to LL-2 at $E_{\rm F}$ (full red line), $I_{\rm stab}=1$\,nA, $V_{\rm stab}=-250$\,mV. The crossing point of the first charging line of LL-3  with the last charging line belonging to LL-2 is marked (red circle) via extrapolation of the two charging lines (dashed red lines).}
    \label{FigS1b}
\end{figure}

These crucial parameters are determined as follows. The sign of $\Delta V_{\rm sample}$  is given by the experimentally observed  ($|m|>0$)-states of LL0 that appear at lower energy than the ($m=0$)-state, itself identified as the line with strongest $dI/dV$ intensity (Fig.~\ref{Fig1}b, main text). This is only possible, if the ($m=0$)-state is confined in a QD potential maximum implying a hole-type QD (Fig.~\ref{FigS1}a). Additionally, the observation that the strongest $dI/dV$ lines of the $(m=0)$-states  are pinned above $E_{\rm F}$, for LL0, LL+1 and LL+2, is consistent with a hole-type TIQD where the   ($m=0$)-states are separated upwards in energy from the bulk LL$n$  pinned at $E_{\rm F}$  (Fig.~\ref{FigS1}c).

 A more detailed consideration of the negative $V_{\rm gate}$ area of Fig.~\ref{Fig1}a, main text, reveals crossings of charging lines originating from different LL$n$ (Fig.~\ref{Fig2}a, main text, and Fig.~\ref{FigS1b}). The attribution of a charging line to LL$n$ uses its intersection with a LL$n$ LDOS feature at $E_{\rm F}$ ($V_{\rm sample}=0$\,V). Hence, the charging lines on the left in Fig.~\ref{FigS1b}  originate from LL-3. They represent the first few holes of LL-3 that are added to the TIQD. The very first LL-3 hole is marked by a dashed line. Close to the crossing of this charging line with $E_{\rm F}$, LL-3 of the surrounding graphene must be completely occupied with electrons (Fig.~\ref{FigS1}b), since the filling factor of the surrounding graphene must always be larger than the filling factor of a hole-type TIQD. Either, $E_{\rm F}$ of the surrounding graphene is in the gap between LL-3 and LL-2 (Fig.~\ref{FigS1}b) or it is at states of the rim of LL-3 that are known to be localized.\cite{Joynt1984,Ando1984} Both situations provide an insulating barrier for the confined charge carriers in the TIQD, such that screening of the added charge is strongly suppressed. Consequently, a strong change of the DOS by charging the TIQD results in a bright charging line.

 The charging lines appearing on the right of Fig.~\ref{FigS1b} belong to the last holes from  LL-2 that are charged into the TIQD. They exhibit a steeper slope since these states are, on average, located further away from the capacitive center of the tip (Fig.~\ref{FigS1}). In the surrounding bulk, $E_{\rm F}$  must be located within LL-2, again since the filling factor of the bulk must be larger than the local filling factor of a hole-type TIQD (Fig.~\ref{FigS1}c).  

 The crossing point of two charging lines from LL-3 and LL-2 implies that QD states from both LLs are at $E_{\rm F}$ simultaneously. This is naturally realized by a ring like charge distribution with occupied hole states including the first state from LL-3 in a central disk and occupied hole states only from LL-2 in an annulus around the disk (Fig.~\ref{FigS1}d). Such configuration is quite usual for QDs in $B$ field and sometimes called a wedding cake.\cite{Gutirrez2018} The wedding cake scenario is also found in our Poisson simulations (not shown).
The situation of Fig.~\ref{FigS1}d can be used to estimate the TIQD depth. Since LL-2 is at $E_{\rm F}$ in the surrounding bulk, the QD depth approximately equals the known energy gap between LL-2 and LL-3 of 30.5\,meV (eq.~\ref{relLandau}). Here, we ignore the finite energetic width of the bulk Landau levels since not being a dominant error. For the transition from LL-1 to LL-2, a crossing of charging lines is also observed (Fig.~\ref{Fig2}a, main text) implying a TIQD depth at this point of 38.9\,meV. From these two crossing points at two distinct pairs of $V_{\rm sample}$ and $V_{\rm gate}$, we eventually determine $\Delta V_{\rm gate}$ and $\Delta V_{\rm sample}$.\\
\begin{figure}
    \centering
    \includegraphics[scale=0.59]{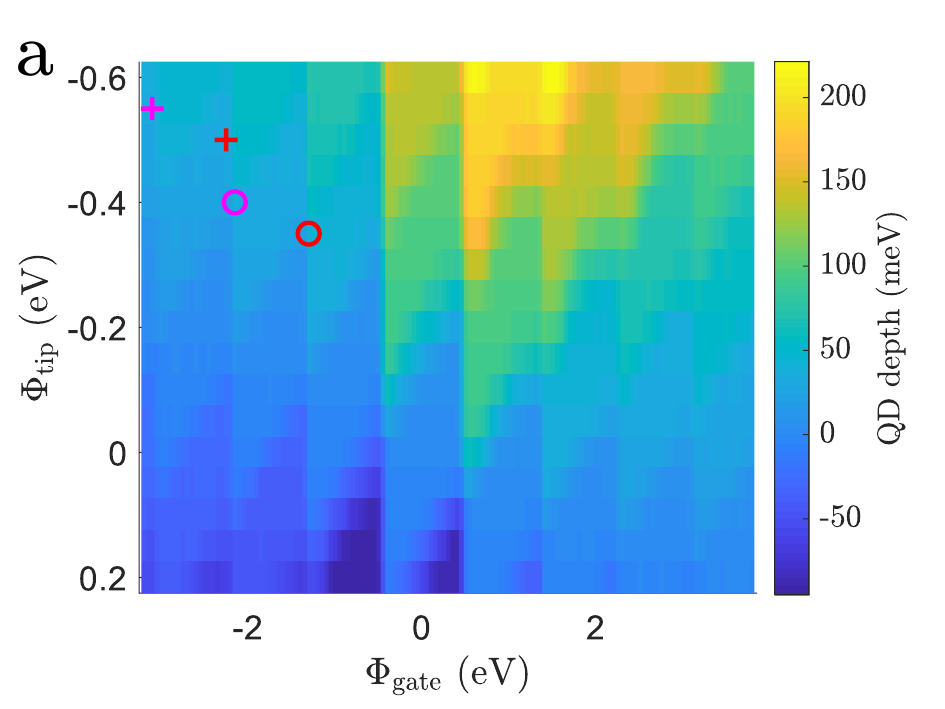}
    \includegraphics[scale=0.59]{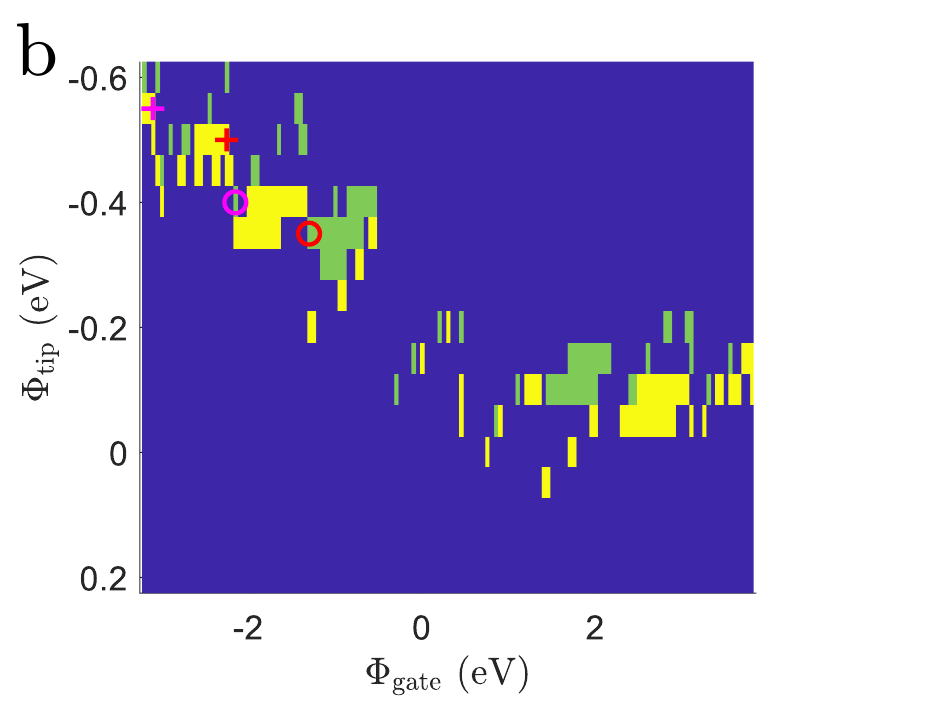}
    \caption{
    (a) Potential depth of the hole-type TIQD for varying external potentials $\Phi_{\rm gate}$ and $\Phi_{\rm tip}$, $r_{\rm tip}=25$\,nm.
    (b) Selection of the potential depth values from a that are in accordance with the crossing points of charging lines in the experiment.
    green: TIQD depth $= 39.8\pm 2.5$\,meV, yellow: TIQD depth $= 30.5\pm 2.5$\,meV, blue: all other TIQD depths. Pairs of
    circle and cross of the same color are separated by $e\delta V_{\rm gate}= 0.95$\,eV along $\Phi_{\rm gate}$ and by $e\delta V_{\rm sample}=0.15$\,eV along $\Phi_{\rm tip}$, such as the two crossing points of charging lines in Fig.~\ref{FigS1b} and Fig.~\ref{Fig2}a, main text. Only two pairs are found to match the required conditions that the circle is on a green area, while the corresponding cross is on a yellow area.
}
    \label{FigS2}
\end{figure}

Practically, we firstly measure the experimental voltage differences between the two crossing points, $\delta V_{\rm gate}=0.95$\,V  in $V_{\rm gate}$ direction and $\delta V_{\rm sample}=0.15$\,V in $V_{\rm sample}$ direction.
Then, we determine the depth of the TIQD potential from the Poisson simulations at varying $\Phi_{\rm tip}$ and $\Phi_{\rm gate}$ (Fig.~\ref{FigS2}a) using circular symmetric coordinates (Fig.~\ref{FigS4}a). Afterwards, we select all ($\Phi_{\rm tip}$, $\Phi_{\rm gate}$) that exhibit the potential depths as present during the crossing points in the experiment (color code in Fig.~\ref{FigS2}b).
Subsequently, we find pairs of
($\Phi_{\rm gate}$, $\Phi_{\rm tip}$) that
feature the two TIQD depths at the two crossing points (38.9\,meV, 30.5\,meV) and, at the same time, the energetic distances in $\Phi_{\rm gate}$ and $\Phi_{\rm tip}$ that are identical to the voltage distances between the two crossing points ($\delta V_{\rm gate}=0.95$\,V, $\delta V_{\rm sample}=0.15$\,V). The found pairs are marked as symbols of the same color in Fig.~\ref{FigS2}.

This still leaves us with two possibilities.
To select the correct one, we compare the two pairs of two ($\Phi_{\rm tip}, \Phi_{\rm gate}$) with the respective two  ($V_{\rm gate}$, $V_{\rm sample}$) of the two crossing points to determine their offsets, $\Delta V_{\rm sample}$ and $\Delta V_{\rm gate}$. Then, we compare the resulting calculated LDOS($V_{\rm gate}$, $V_{\rm sample}$) for both cases with the measured $dI/dV_{\rm sample}(V_{\rm gate},\,V_{\rm sample})$ (Fig. \ref{FigS3}).
This leads to a straightforward selection of the the red pair within Fig.~\ref{FigS2} corresponding to $\Delta V_{\rm gate}= -200\pm 50$\,mV and $\Delta V_{\rm sample}= -180\pm 50$\,mV. The relatively large error of these values results from the selected step size of 100\,meV in $\Phi_{\rm tip}$ and $\Phi_{\rm gate}$ within the Poisson simulations.
%
\begin{figure}
    \centering
    \includegraphics[scale=0.59]{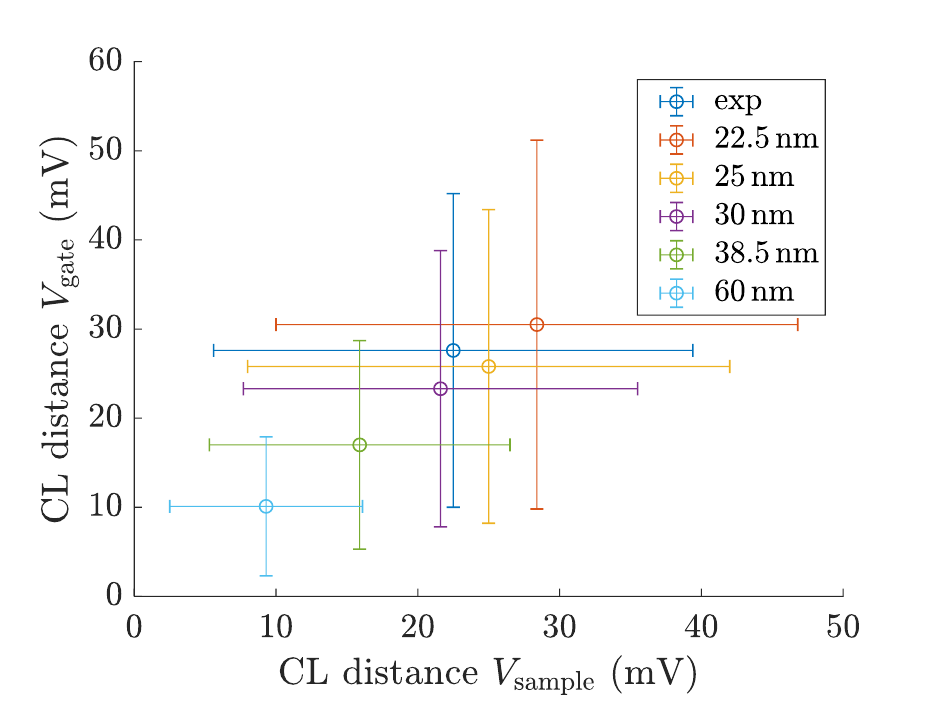}
    \caption{
    {Comparison of the averaged distance between charging lines in the experiment (dark blue) with the ones deduced from the Poisson simulations at different $r_{\rm tip}$ as labelled, $\Delta V_{\rm gate}=-200\pm 50$\,mV, $\Delta V_{\rm sample}=-180\pm 50$\,mV. The error bars of the experiment result from the variance of the averages from different regularly spaced groups of charging lines (see text). The error bars of the simulations result from the variance in $\Delta\Phi_{\rm gate}/\Delta Q_{\rm QD}$ and $ \Delta\Phi_{\rm tip}/\Delta Q_{\rm QD}$}, respectively, within the simulation range of  $\Phi_{\rm tip} \in{[-0.6,0]}\,$eV and $\Phi_{\rm gate} \in{[-3.2,0]}\,$eV. Note that the error bars indicate the same variance in experiment and simulation, but do not provide the statistical uncertainty of the mean values.
}
    \label{FigS5}
\end{figure}

The remaining fit parameter $r_{\rm tip}$ is deduced from the distance of the charging lines in the experiment by comparison with the Poisson simulations.
In the Poisson simulations, we determine the additional charge within the TIQD, $\Delta Q_{\rm QD}$, that is caused by a potential change $\Delta\Phi_{\rm tip}$ in $\Phi_{\rm tip}$ direction or $\Delta\Phi_{\rm gate}$ in $\Phi_{\rm gate}$ direction.
Note that $e\Delta Q_{\rm QD}/\Delta \Phi_{\rm gate}$ is directly the capacitance of the TIQD with respect to the gate as often used for analyzing quantum dots in transport experiments.\cite{Ihn}
The total charge $Q_{\rm QD}$ within the TIQD is calculated by spatially integrating the confined charge carrier density up to the edge of the TIQD. The edge separates the TIQD from the surrounding bulk with constant filling factor and, hence, can include an outer insulating (incompressible) ring of the TIQD where the potential is still changing (blue areas in Fig.~\ref{FigS1}b).
Eventually, we compare $\Delta Q_{\rm QD}/\Delta \Phi_{\rm gate}$ and $\Delta Q_{\rm QD}/\Delta \Phi_{\rm tip}$ for various $r_{\rm tip}$ with the experimental number of charging lines per voltage (Fig.~\ref{FigS5}).
For this purpose, we select groups of charging lines with regular voltage distances implying only minor contributions from orbital energy, i.e., from the confinement energy neglected in the Poisson simulations. 
We determine their average distance and use the average of all such groups for $V_{\rm gate}<0$\,V and $V_{\rm sample}>0$\,V.
This voltage area is selected since the simulated LDOS data at $V_{\rm gate}<0$\,V matches the experiment favorably (Fig.~\ref{FigS3}) and since each charging line should only contribute once. The preselection of groups of regular charging lines also deals with the fact that some of the charging lines might not be visible due to imperfect confinement at $E_{\rm F}$ or strong screening from the surrounding graphene.

Practically, we firstly estimate $r_{\rm tip}$ by adapting the ratio of $(\Delta Q_{\rm QD}/\Delta\Phi_{\rm gate})/(\Delta Q_{\rm QD}/\Delta\Phi_{\rm tip})$ to the corresponding ratio of the experiment (slope of the charging lines) and latter refine via the agreement of absolute values of $\Delta Q_{\rm QD}/\Delta\Phi_{\rm gate}$ and $\Delta Q_{\rm QD}/\Delta\Phi_{\rm tip}$ with the experimental ones. The parameter $r_{\rm tip}$ is varied until the absolute values fit  favorably
resulting in $r_{\rm tip}=25$\,nm (Fig.~\ref{FigS5}).
We finally check for consistency by repeating the determination of $\Delta V_{\rm sample}$ and $\Delta V_{\rm gate}$ with the found $r_{\rm tip}$ (Fig.~\ref{FigS2}). However, we find that these two values barely depend on $r_{\rm tip}$.

In principle, the Poisson simulations also reveal the charge in the TIQD for each $(V_{\rm gate}, V_{\rm sample})$. Hence, one could add lines to Fig.~\ref{FigS3}b--c at integer multiples of $e$ in the TIQD in order to also reproduce the charging lines. We crosschecked that this partially matches the experiments, but generally would overemphasize the accuracy of our model.

\subsection{Poisson Simulation Including the Lateral Interface}
\label{sec:PoissonlatInt}
For determining the full potential profile across the interface in presence of the TIQD, we employ 2D Cartesian coordinates that neglect the direction along the interface (Fig.~\ref{FigS4}a). We crosschecked that this 2D restriction gives the same result as the 3D calculation of the TIQD using cylindrical symmetry far away from the interface.
The potential profile is determined using the geometry as depicted in Fig.~\ref{FigS4}a and employs the determined $r_{\rm tip}$, $\Delta V_{\rm gate}$, and $\Delta V_{\rm sample}$. We perform simulations for multiple positions of the tip $x_{\rm tip}$ while varying $\Phi_{\rm tip}$ and $\Phi_{\rm gate}$ (e.g. Fig.~\ref{Fig2}c, main text).

\subsection{Tight Binding Model of the Tip Induced Quantum Dot}
\label{sec:TB_TIQD}
\begin{figure}[h]
	\centering	
	\begin{tikzpicture}	
		\node[inner sep=0pt] (russell) at (-0.51,0)   {\includegraphics[trim=0 0 0 0 ,clip,width=0.34\textwidth]{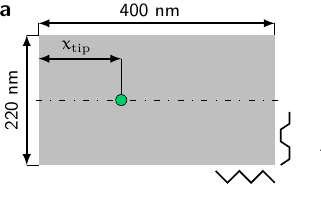} };
        \node[inner sep=0pt] (russell) at (0,-6)   {\includegraphics[trim=0 0 0 0 ,clip,width=0.45\textwidth]{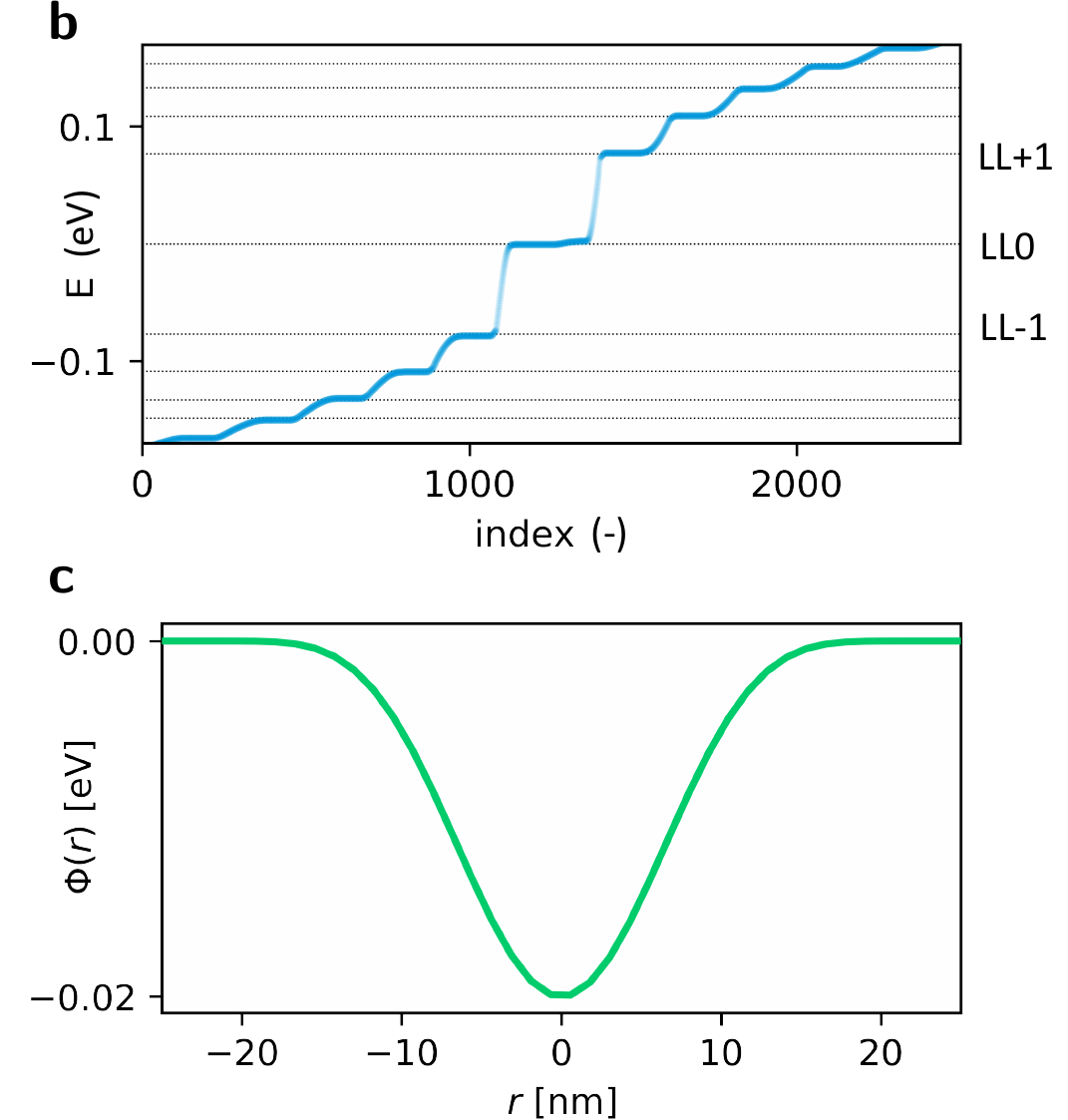} };			
        \node[inner sep=0pt] (russell) at (-1.58,-3.3)   {\includegraphics[trim=0 0 0 0 ,clip,width=0.15\textwidth]{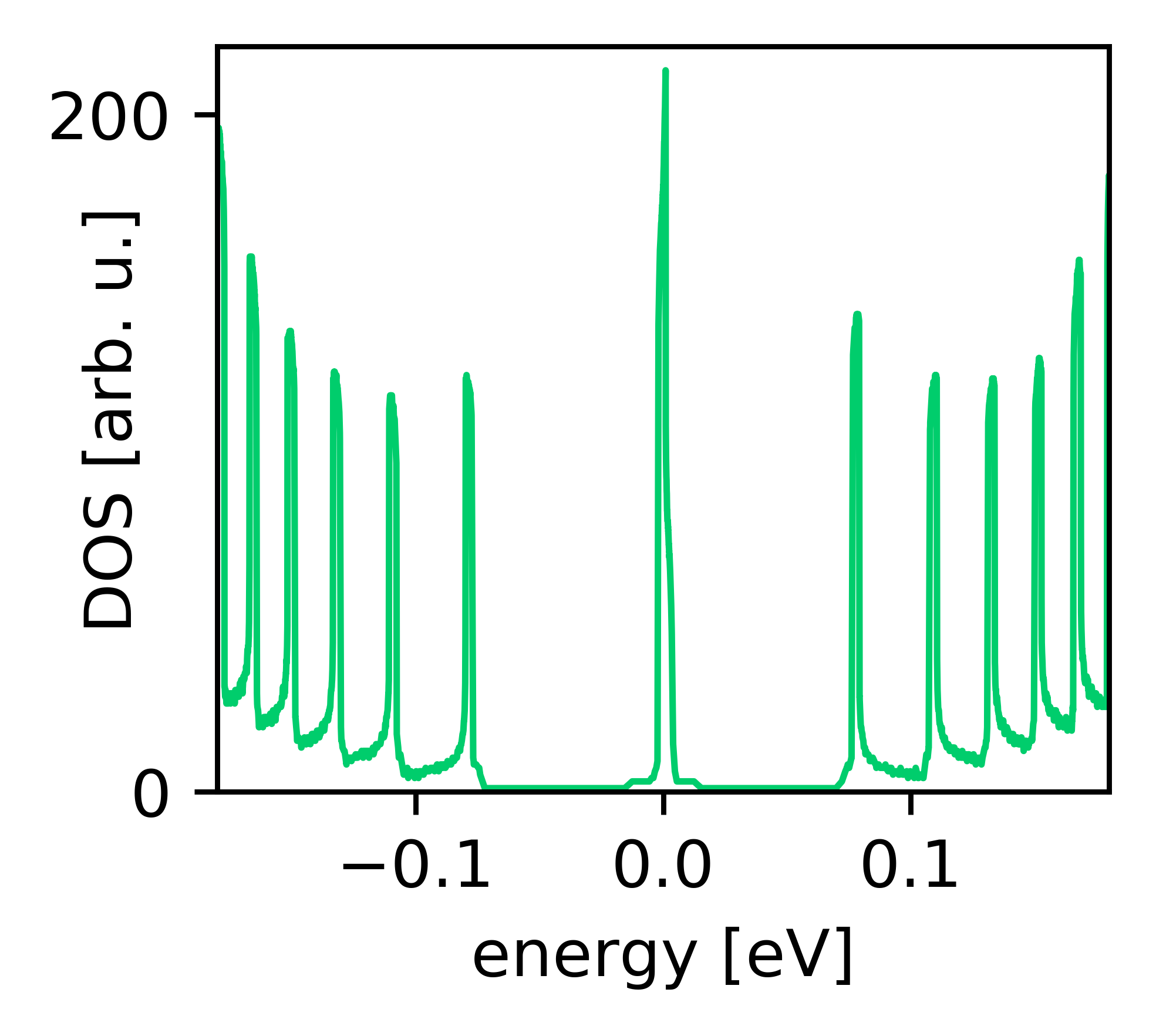} };		
        \draw[white,very thick,fill=white] (-0.2,-9.85) circle (0.2);
        \draw[white,very thick,fill=white] (0.,-9.85) circle (0.2);
        \draw[white,very thick,fill=white] (0.2,-9.85) circle (0.2);
        \draw[white,very thick,fill=white] (0.4,-9.85) circle (0.2);
        \draw[white,very thick,fill=white] (0.6,-9.85) circle (0.2);
        \draw[white,very thick,fill=white] (0.8,-9.85) circle (0.2);
        \node[inner sep=0pt] (russell) at (-0.31,-8.5)   {\includegraphics[trim=22 0 15 0 ,clip,width=0.4\textwidth]{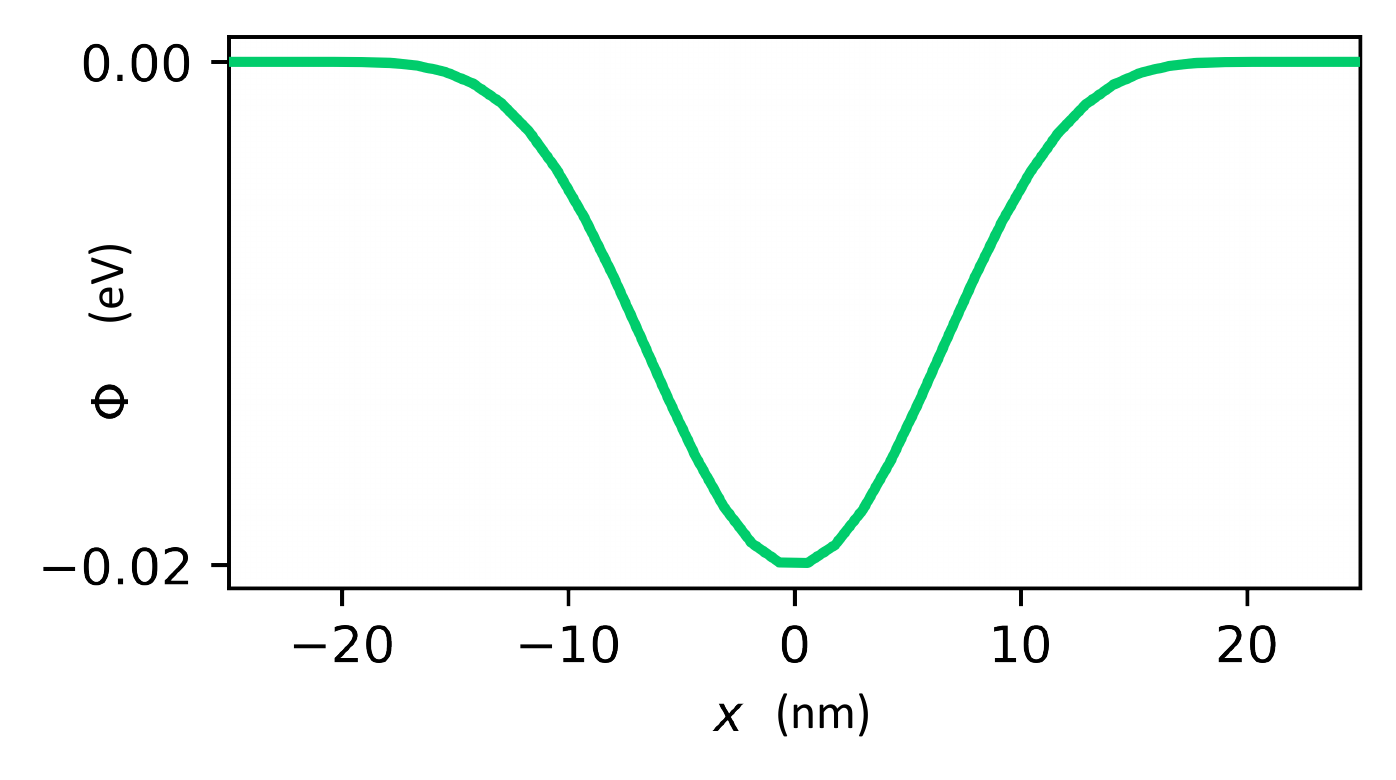} };	
	\end{tikzpicture}
	\caption{
	    (a) Top view of the simulated graphene flake 
	    with marked edge types and center position of the TIQD potential (green dot). (b) Eigen-energies of the  graphene at $B=7$\,T. Horizontal dotted lines mark the analytic LL energies (\cref{relLandau}). Inset: resulting density of states. (c) Exemplary electrostatic potential of the TIQD that is imprinted on the graphene sheet by  the STM tip (\cref{e:potential_dumb1}, \cref{e:potential_dumb}).
		}
	\label{fig:LL_intro}	
\end{figure}
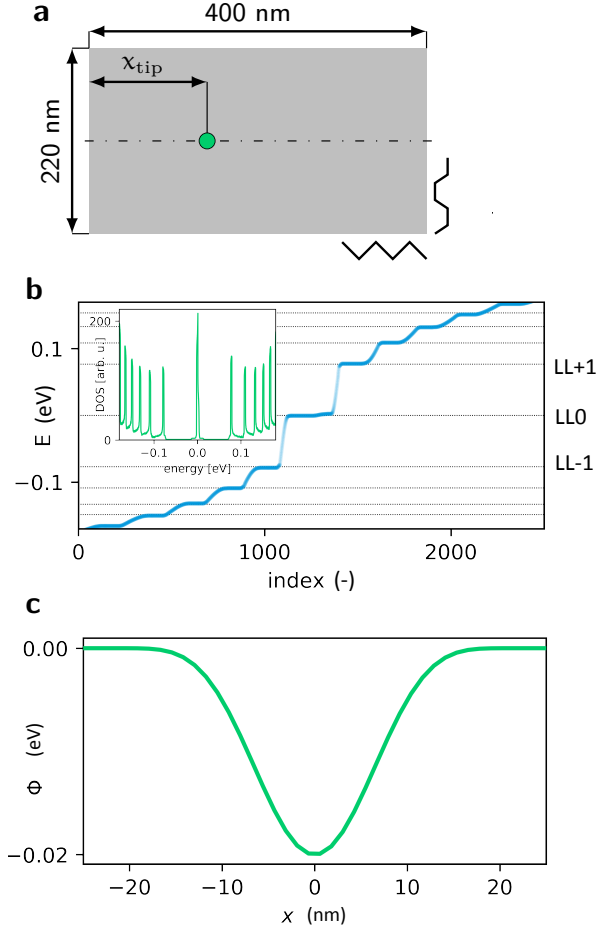

Before discussing the TB simulations of the lateral interface with TIQD, we describe the TB results for the TIQD without interface. Fig.~\ref{fig:LL_intro}a sketches the investigated large graphene flake with chosen zigzag and armchair edges. A center position of the TIQD is marked at $x_{\rm tip}$ in horizontal direction. In vertical direction, the TIQD is always centered in the middle of the flake. Fig.~\ref{fig:LL_intro}b shows the resulting energy spectrum with 2500 states for the graphene rectangle without TIQD at $B=7$\,T. The corresponding density of states (DOS) (inset of Fig.~\ref{fig:LL_intro}b) reveals pronounced Landau quantization with peak energies according to eq.~(\ref{relLandau}). We adapted the energy scale to account for the slightly different $v_{\rm F}$ deduced from the LL$n$ energy distance in experiment ($1.0 \cdot 10^6$\,m/s) and
resulting from our $3^{\text{rd}}$-nearest-neighbour TB parameters ($0.8\cdot 10^6$\,m/s).

Next, we add the potential of the TIQD  $\Phi_{\mathrm{TIQD}}(\mathbf{x})$ with center at the marked position $x_{\rm tip}$. We use a fit function found previously \cite{Freitag2018,PhysRevB.102.155430,Schnez2008} reading
	\begin{align}
	    \Phi_{\mathrm{TIQD}}(\mathbf{x}) =
	    \begin{cases}
	        -V_0 \cdot \mathrm{cos}\big(\frac{\pi}{2\alpha} |\mathbf{x}| \big)^5 &, |\mathbf{x}| < \alpha
	        \\
	        0 &, |\mathbf{x}| \geq \alpha
	    \end{cases}
	    \label{e:potential_dumb1}
	    \\
	    \text{with}~~ \alpha = 2309\cdot |V_0| \sqrt{1 + \sqrt{\frac{0.4}{0.005+|V_0|}}}.
	\label{e:potential_dumb}
	\end{align}
with $\mathbf{x}$ being the 2D position on the flake with respect to $\mathbf{x}_{\rm tip}$. The parameter $\alpha$ is taken in \AA, while $|V_0|$ is taken in eV. For demonstration in Fig.~\ref{e:potential_dumb}c, we use a potential depth $V_0 = 0.02$\,eV, but later $V_0$ and $\alpha$ are fit parameters that are optimized to reproduce the potential profiles from the Poisson simulations.

\begin{figure}[H]
\centering
\includegraphics[width=0.235\textwidth]{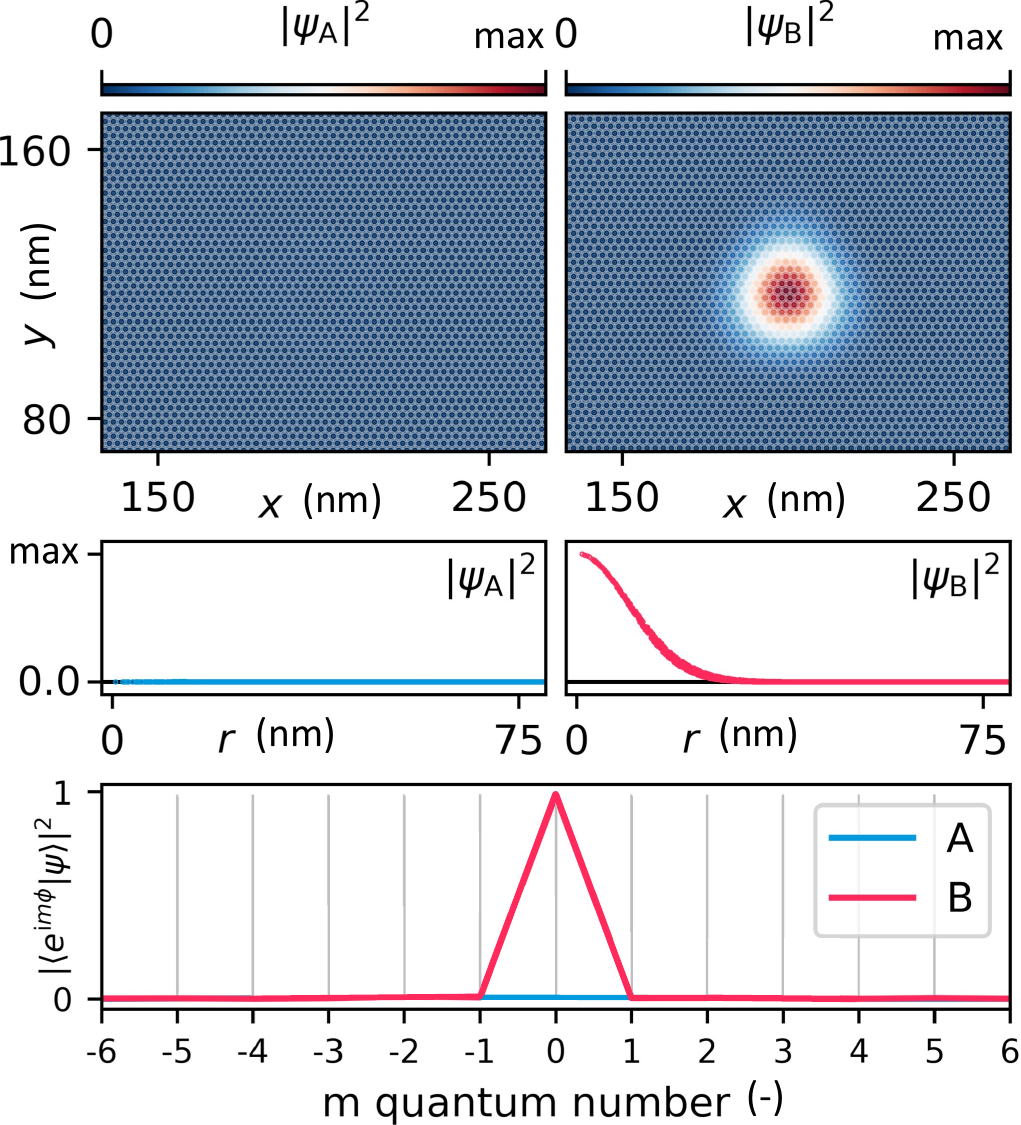}
\includegraphics[width=0.235\textwidth]{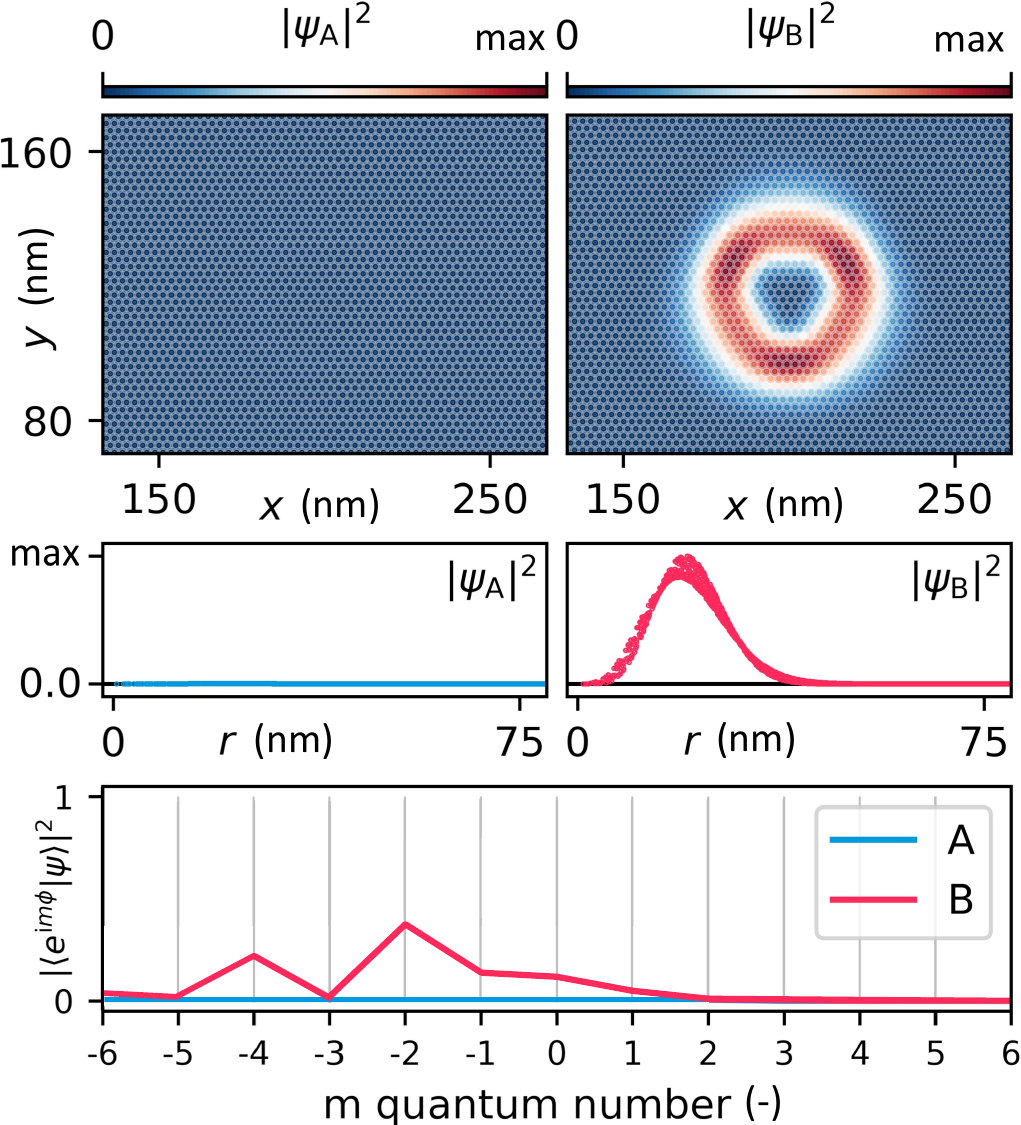}
\includegraphics[width=0.235\textwidth]{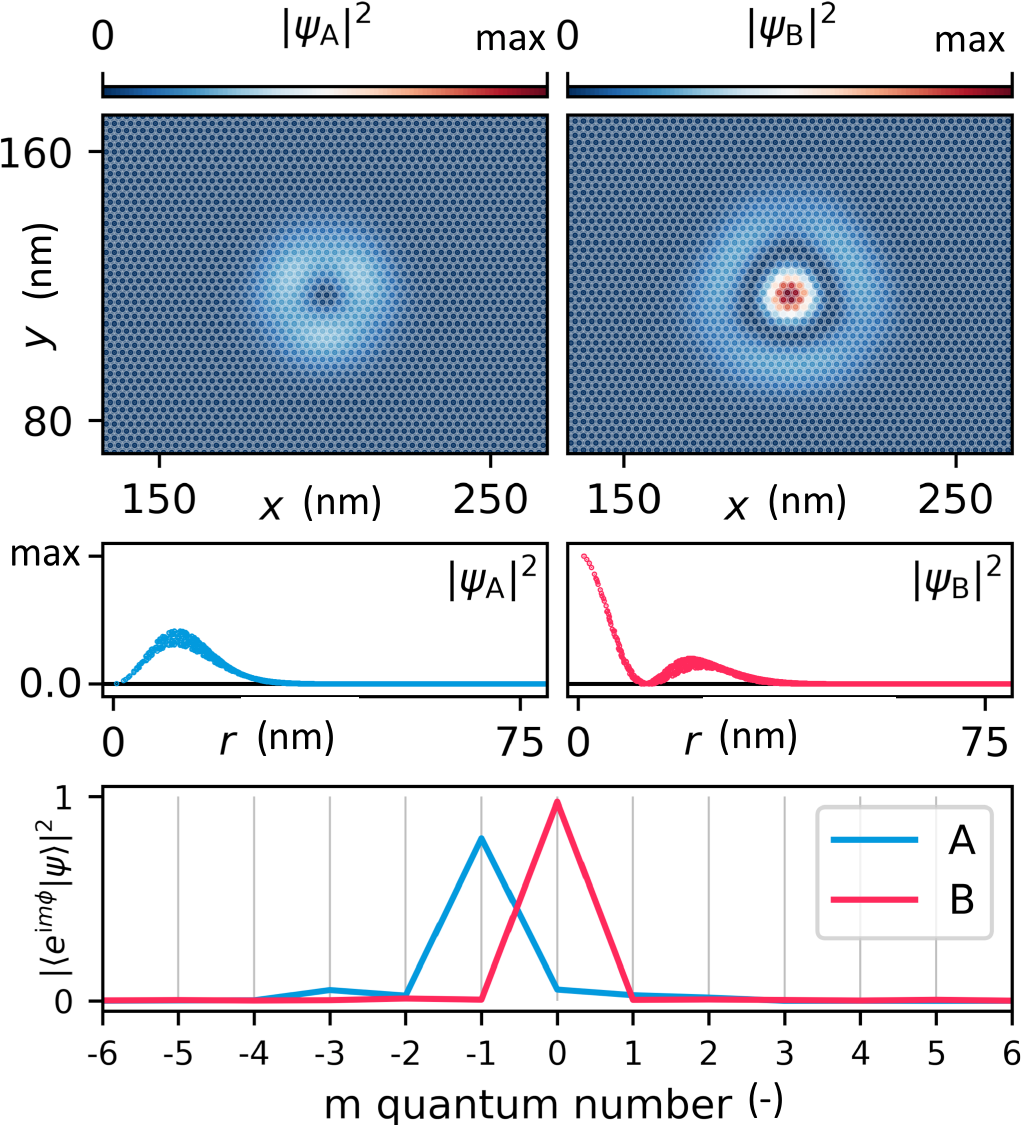}
\includegraphics[width=0.235\textwidth]{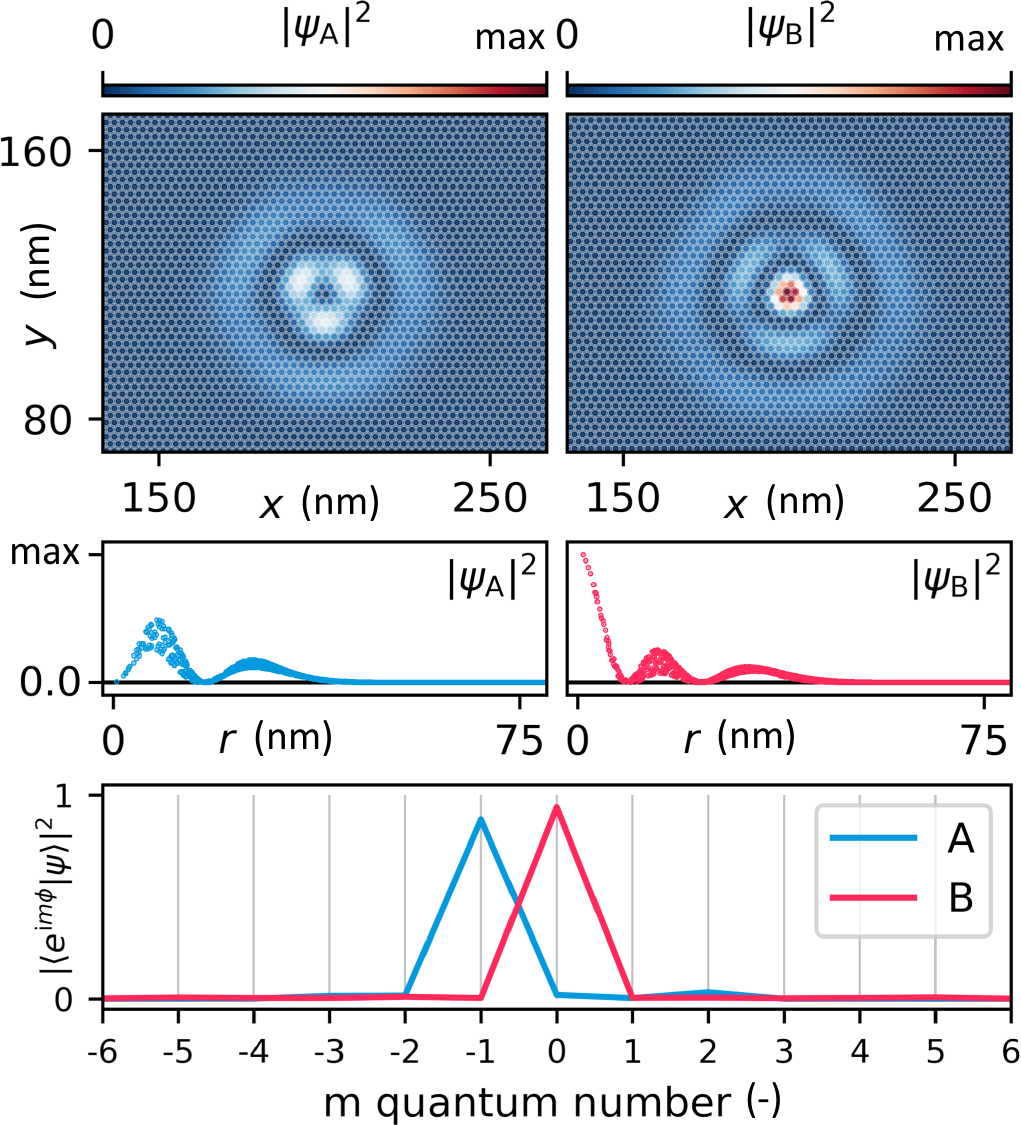}
\begin{picture}(1,1)
\put(-122,270){\large{\textsf{\textbf{{a}}}}}
\put(4,270){\large{\textsf{\textbf{{b}}}}}
\put(-122,136){\large{\textsf{\textbf{{c}}}}}
\put(4,136){\large{\textsf{\textbf{{d}}}}}
\end{picture}
\caption{(a)--(d) Analysis of four TIQD states, namely (a) LL0 \#0, (b) LL0 \#4, (c) LL1 \#2 , and (d) LL2 \#4. The number after \# counts the states at energies below the corresponding bulk LL$n$ starting with the one at lowest energy (dubbed \#0). Upper row:
2D color plots of the probability densities $|\psi_{A/B}|^2$ for each sublattice A and B separately. Middle row: radial density of $|\psi_{A/B}|^2$ for both sublattice contributions. Lower row:
Overlap  with azimuthal test functions  $\ket{\psi_{{\rm test},m}}=|\mathrm{e}^{\text{i}m \varphi} \rangle$ according to $|\langle{\psi_{{\rm test},m}}\ket{\psi_{\rm A/B}}_\varphi|^2$ integrated along the azimuthal angle $\varphi$ within an annulus around the global density maximum in radial direction.
}
\label{fig:QDOT_quantum_numbers}
\end{figure}
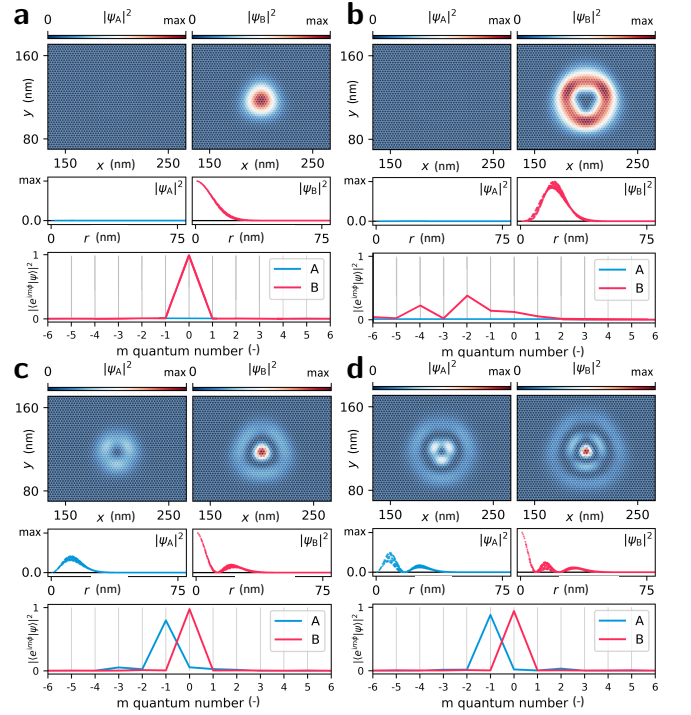

The resulting energy spectrum of the flake with TIQD consists of $\sim 2000$ eigenstates where the localized quantum dot states are energetically separated from the LL$n$ energies (\cref{relLandau}). Fig.~\ref{fig:QDOT_quantum_numbers} displays some of these states. They showcase the typical sublattice-dependent structure that appears for graphene LLs at each valley $K$ or $K'$.\cite{CastroNeto2009}
The index $n$ describes the conventional LL wave functions
and, hence, differs by one between the two sublattice components reading
\begin{equation}
	\ket{\psi^{K}_n} = \left(\begin{array}{c}\ket{\phi _{|n|-1}}\\\ket{\phi _{|n|}}\end{array}\right),\qquad\ket{\psi^{K'}_n} = \left(\begin{array}{c}\ket{\phi _{|n|}}\\\ket{\phi _{|n|-1}}\end{array}\right)\label{valley}
\end{equation}
with $\ket{\phi _{|n|}}$ being the classical LL wave functions. For $n \equiv 0$, the other component 
$\ket{\phi _{|n|-1}}$ vanishes. 

To analyze the calculated confined states within the TIQD, we energetically separate these sublattice structures by numerically breaking valley degeneracy with a  potential in sublattice space reading $10\, \text{$\mu$eV} \cdot \sigma_z$.\cite{Freitag2018, Freitag2016}.

For quantum dots with a spherical infinite mass boundary or with zig-zag boundaries of the continuum Dirac-Weyl Hamiltonian, one gets well-defined radial and azimuthal quantum numbers $n_r\in \mathbb{N}_0$ and $m\in \mathbb{Z}$, respectively, that are related to the LL index $n$ via
 \cite{Schnez2008}	

 \begin{align}
		n&=n_r+\frac{m+|m|}{2} 		\label{e:mapping_PeetersIMBC}
		\end{align}
for infinte mass boundary conditions or via		\begin{align}		
		n&=n_r+\frac{m+|m|}{2} - \Theta(m) \label{e:mapping_PeetersZZBC}
	\end{align}
for zig-zag boundary conditions with Heavyside function $\Theta$.	
Note that each state that belongs to a particular LL$n$ is uniquely defined by the index $m$.

Albeit this model is not entirely applicable to our boundary conditions, we analyze the TIQD states accordingly (Fig.~\ref{fig:QDOT_quantum_numbers}).
The number of radial nodes is easily determined by inspecting the radial density distribution (middle row of Fig.~\ref{fig:QDOT_quantum_numbers}a--d). The angular quantum number is more tricky. We first have to account for the Bloch phase $\text{e}^{\text{i}\mathbf{k} \cdot \mathbf{r}}$ depending on the valley index ($\mathbf{K}=( 4\pi/(3a), 0)$, $\mathbf{K'}=-\mathbf{K}$). Since our confinement potential is smooth, valley is still a good quantum number as verified by inspecting the localization of Husimi distributions in reciprocal space (not shown). After removing the Bloch phase, we select a slim annular region around the global, radial maximum of each state. 
After renormalizing within this area, we calculate the overlap integrals $|\langle \psi_{{\rm test},m} | \psi_{A/B} \rangle_\varphi|^2$ with test functions of the form $\langle \mathbf{r} | \psi_{{\rm test},m}  \rangle = \text{e}^{\text{i} m \varphi(\mathbf{r})}$, $m \in [-10,10]$, where $\varphi$ is the azimuthal angle 
calculated with respect to the TIQD center. A large overlap as in Figs.~\ref{fig:QDOT_quantum_numbers}a, c, d, bottom rows, indicates a well-defined  quantum number $m$ of the corresponding TIQD states.
This is, however, not always the case as, e.g., in Fig.~\ref{fig:QDOT_quantum_numbers}b, bottom row, exhibiting two relevant $m$ overlaps. The discrepancy is likely due to effects of trigonal warping that are additionally enhanced by the applied artificial enlargement of the unit cell.
Moreover, we find that the conditions of
\cref{e:mapping_PeetersIMBC,e:mapping_PeetersZZBC} do partially not hold.
Nevertheless, for the sake of simplicity, we dub the states found in the experimental data as $m$ states of a particular LL$n$ in order to stress the central antinode of the $m\simeq 0$ state that is most strongly visible in the STM data and is also found consistently for different LL$n$  in the TB simulations (e.g.  Figs.~\ref{fig:QDOT_quantum_numbers}a, c, d, right column, top and bottom row).

\begin{figure}[t]
\centering
\begin{tikzpicture}
    \node[inner sep=0pt] (russell) at (0,0)   {	\includegraphics[trim= 0 0 0 0, clip,width=0.49\textwidth]{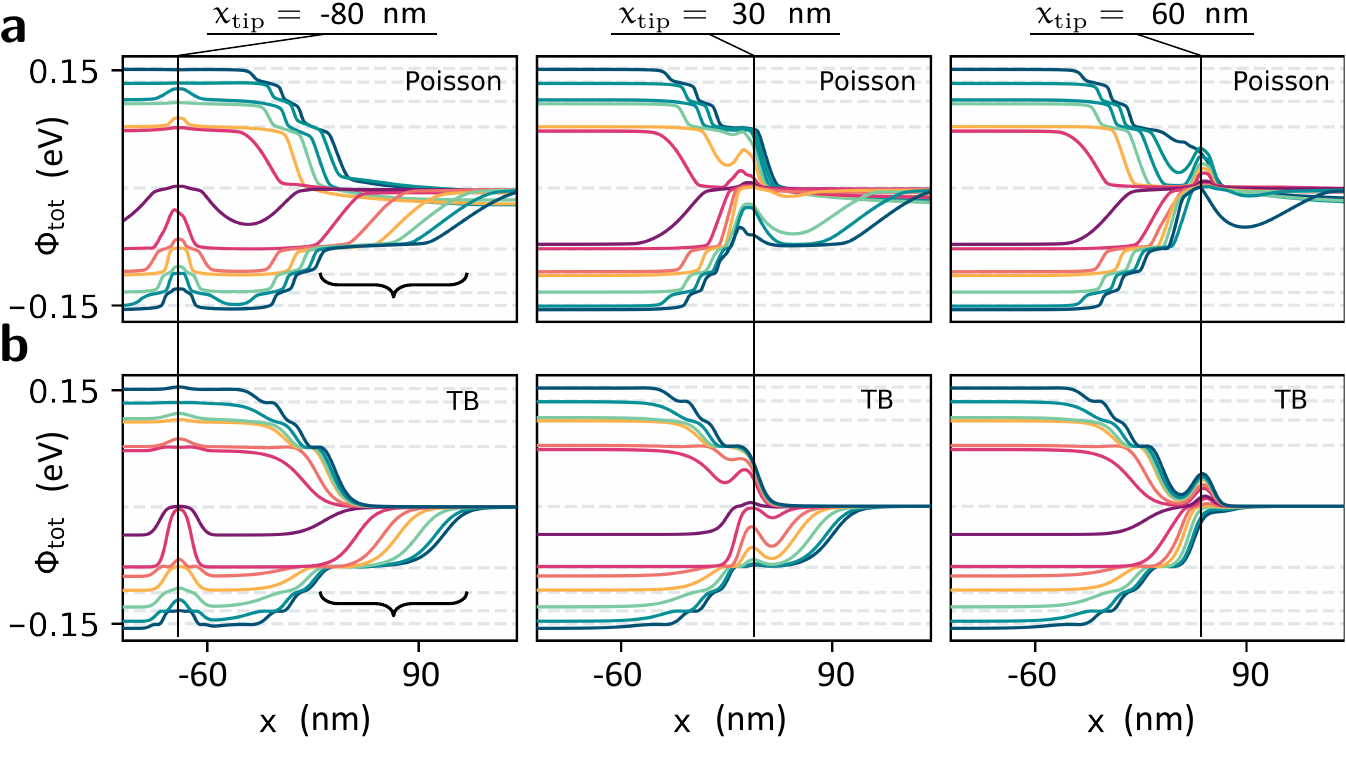}};		
\end{tikzpicture}
\caption{
    (a) Electrostatic potentials from the Poisson simulations of the lateral interface with TIQD as displayed along a 1D trajectory perpendicular to the interface. Different $V_{\rm gate}$ are applied in each sub-figure from $+3.5$\,V (bottom curve) to $-3$V (top curve) with increments of $-0.5$\,V. Each sub-figure displays a different center position of the TIQD $x_{\rm tip}$ as marked by the vertical black line, $V_{\rm sample}=0$\,V (b) Linecut through the 2D analytic potential resulting from a fit of the curves in a via eq.~(\ref{e:tot_potential})--(\ref{e:quench_fun}) with fit parameter $\eta$, that is changed only for different $V_{\rm gate}$, and quench function parameters $a_i$, $b_i$, $c_i$, that are adapted for each curve. The line cut is along the central horizontal line of the 2D area (dashed-dotted line in Fig.~\ref{fig:LL_intro}a). The brackets in the left row of images mark a lateral shift of the potential step at the interface by the TIQD that is rather well reproduced by our fit functions.
}
\label{fig:1d_2d_pot}
\end{figure}

\subsection{Transferring the 1D Poisson Simulation to the 2D Potential of the Tight Binding Model}
\label{sec:PoissontoTB}
Next we use the results from the 1D Poisson simulations to obtain a 2D potential as base for the TB model. For ease of implementation, we use analytic fit functions.

We parameterize the total potential as
\begin{align}
 	\Phi_{\mathrm{tot}}(\mathbf{x}) = q\kl{ \lambda(\eta, x_{\rm tip})\cdot  \Phi_{\mathrm{TIQD}}^{(\mathrm{TB})}(\mathbf{x})  + \eta\cdot \Phi_{\mathrm{int}}^{(\mathrm{TB})}(\mathbf{x}) }.
\label{e:tot_potential}
\end{align}

The argument of the quench function $q(\Phi)$ is a superposition of the TIQD potential and the lateral interface potential. The TIQD potential reads (eq.~\ref{e:potential_dumb1})
	\begin{align}
		&\Phi_{\mathrm{TIQD}}^{(\mathrm{TB})} (\mathbf{x}, \mathbf{x}_{\mathrm{tip}})= 
		\begin{cases}
			\mathrm{cos}^5\kl{\frac{\pi}{2} \frac{ |\mathbf{x}\shortminus\mathbf{x}_{\mathrm{tip}}|}{31.95} },     & |\mathbf{x}\shortminus\mathbf{x}_{\mathrm{tip}}| < 31.95
            \\ 0, & |\mathbf{x}\shortminus\mathbf{x}_{\mathrm{tip}}| \geq 31.95.
        \end{cases}\\
		&~\nonumber\\
		&\mathrm{It~is~weighted~by~prefactor~\lambda~ that~depends~on~the~potential} \nonumber\\
		&\mathrm{drop~across~the~ interface~parametrized~ by}~\eta~\mathrm{and~on}~x_{\rm tip}: \nonumber\\
		&~\nonumber\\
		&{\scriptstyle \lambda(\eta, x_{\mathrm{tip}} ) = \shortminus0.06 \shortminus \frac{0.0005~ \mathcal{F}(x_{\mathrm{tip}}, 2200, 30)}{(\eta+0.19)^2+0.005} +\Big(0.06\shortminus0.08\kl{\eta+0.42}\Big) \times }\nonumber\\
		&{\scriptstyle ~~~~~~~~~~~~~~\Big(
		1\shortminus\mathcal{F}\kl{x_{\mathrm{tip}},2700,20}
		\Big) \mathcal{F}\kl{\eta, 0.2+\kl{\frac{x_{\mathrm{tip}} \shortminus1300}{180}}^8, 0.01} \label{eq:lambda}}
	\end{align}	
with Fermi function $\mathcal{F}(x, \mu, \sigma) = \frac{1}{\mathrm{e}^{(x-\mu)/\sigma} +1}$.
The tip position $x_{\rm tip}$ is given in units of \AA~and the resulting $\lambda(\eta, x_{\rm tip})$ in units of eV.

The one-dimensional potential across the lateral interface is modelled by
	\begin{align}
	    &~\nonumber\\
	    &{\scriptstyle\Phi_{\mathrm{int}}^{(\mathrm{TB})} (x) = \mathcal{F}\kl{x, 2216,14} + \Theta \kl{\shortminus \eta} \mathrm{max}\kl{\shortminus0.13,\eta} \times} \nonumber\\
	    &{\scriptstyle~~~~~~~~~\mathcal{F}\Big(x,2216\shortminus 2300\eta~\mathcal{F}\kl{x_{\mathrm{tip}},2700,20}, 14\Big)  \mathcal{F}\kl{\shortminus x,\shortminus2216,14}}
	    &~\nonumber\\
	    \label{e:pn_potential}
	\end{align}
also with energies in eV and position parameters in \AA.
\begin{figure}
\centering
\begin{tikzpicture}
	\node[inner sep=0pt] (russell) at (0,0)   {	\includegraphics[trim= 0 0 0 0, clip,width=0.49\textwidth]{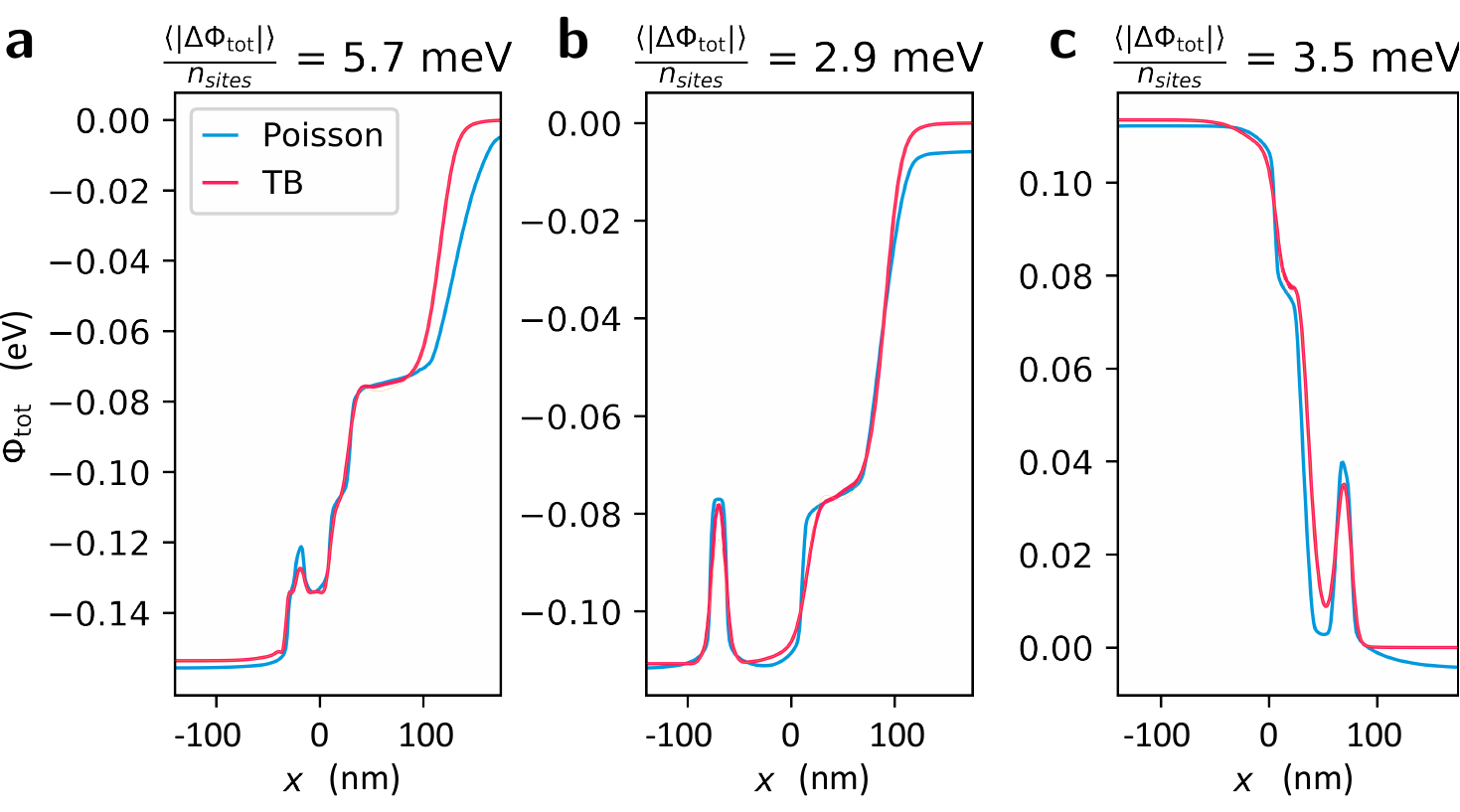}};
\end{tikzpicture}
\caption{(a)--(c) Direct comparison of the electrostatic potentials calculated with the 1D Poisson solver (blue) and the analytic fits used for the 2D TB calculations  (red) for three representative $V_{\text{gate}}$ and $x_{\text{tip}}$: (a) $V_{\text{gate}} = 3.5$\,V ($\eta=\shortminus 0.41$\,eV), $x_{\text{tip}} = -20$\,nm, (b) $V_{\text{gate}} = 2$\,V ($\eta=\shortminus 0.31$\,eV), $x_{\text{tip}} = -80$\,nm, (c) $V_{\text{gate}} = -2$\,V ($\eta= 0.31$\,eV), $x_{\text{tip}} = 70$\,nm. The mean of the absolute difference of the two potentials per TB site $\frac{\langle | \Delta \Phi_{\mathrm{tot}}| \rangle}{n_{\rm sites}}$ is marked on top, $V_{\rm sample}=0$\,V.
$V_{\rm gate}$ is different from $\eta$, since neither $\Phi_{\rm int}^{\rm (TB)}$ (eq.~(\ref{e:pn_potential}))  nor $q(\Phi)$ (eq.~(\ref{e:quench_fun})) are normalized to one as well as due to the implicit use of $\eta$ in eq.~(\ref{eq:lambda}).
}
\label{fig:app_pot_diff}
\end{figure}

To incorporate the flattened regions of the interface potential (compressible stripes) resulting from the Poisson solver (Fig.~\ref{fig:1d_2d_pot}a), 
we adapt a quench function $q(\Phi)$ within eq.~(\ref{e:tot_potential}) that locally modifies the potential values by subtracting Gaussians from the unperturbed weight factor of one,
		 \begin{equation}
		 	q(\Phi) = \Phi \cdot \kl{1-\sum_i a_i \mathcal{G}(\Phi, b_i, c_i) },
			 \label{e:quench_fun}
		 \end{equation}
		 with Gaussians $\mathcal{G}(\Phi, b_i, c_i) = \frac{1}{\sqrt{2\pi c_i^2 }} \text{e}^{ \shortminus \frac{(\Phi-b_i)^2}{2 c_i} }$ along the $\Phi$ direction. The Gaussians are centered at $b_i$, with a standard deviation $c_i$ and a height $a_i$.
		Here, $\Phi$ represents the unquenched potential. The $a_i$, $b_i$, $c_i$ with $i \in$ [LL$\shortminus4$, \dots, LL+4 ] are fit parameters describing the flat potential areas (compressible regions) for each LL$n$.
To compare with the experimental data, we account for $\Delta V_{\rm sample}$ and $\Delta V_{\rm gate}$ by adequate energy shifts.


The resulting potential $\Phi_{\mathrm{tot}}(\mathbf{x})$ reproduces all of the relevant features generated by the Poisson solution (Figs.~\ref{fig:1d_2d_pot}a,b). This includes the variations of depth and lateral size of the TIQD  across the lateral interface  as well as the flat potential regions appearing when LL energies cross $E_{\rm F}$. 
 Even complex features such as a pronounced shift of the interface potential step by the TIQD are rather well reproduced (brackets in Fig.~\ref{fig:1d_2d_pot}, left row).  A quantitative comparison is shown for three examples in Fig.~\ref{fig:app_pot_diff} revealing deviations in the few meV regime that we regard as irrelevant considering the uncertainties of the Poisson simulations (section~\ref{sec:Poissonsim}--\ref{sec:PoissonlatInt}) such as the neglected confinement energies within the TIQD and the assumption of a circular symmetric tip.

Since $\Phi_{\mathrm{tot}}(\mathbf{x})$ is two-dimensional by construction, the 2D shape of the TIQD is apparent while traversing the lateral interface (not shown). It develops from a circular symmetric TIQD on the left of the interface with shape depending on $V_{\text{gate}}$ via an elongated, somewhat skewed TIQD at the interface into an again circular TIQD to the right of the interface, here with depth and shape largely independent of $V_{\text{gate}}$, but depending on $V_{\rm sample}$.

The simulated LDOS($x_{\rm tip}$) without TIQD (blue lines in Fig.~\ref{Fig3}d--f, main text) results from a single TB simulation of LDOS($x$, $y$) with $\lambda =0$ (eq.~\ref{e:tot_potential}) and setting $x=x_{\rm tip}$.
\begin{figure}[thb]
\centering
\begin{tikzpicture}
	\node[inner sep=0pt] (russell) at (0,0)   {	\includegraphics[trim= 0 0 0 0, clip,width=0.49\textwidth]{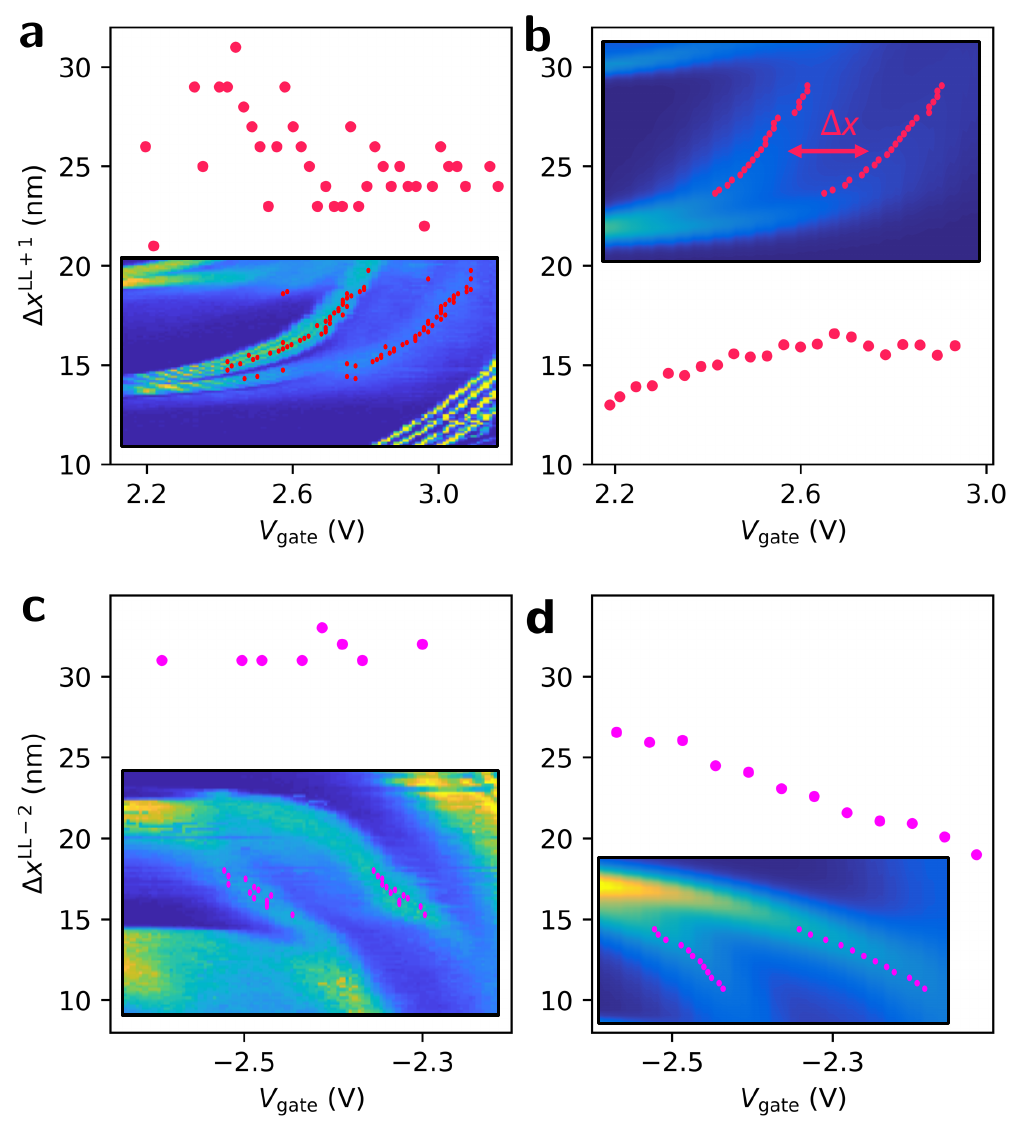}};
\end{tikzpicture}
\caption{
Branching distances $\Delta x$ as deduced from $dI/dV_{\rm sample}(x_{\rm tip}, V_{\rm gate})$ of Fig.~\ref{Fig1}g, main text, and LDOS$(x_{\rm tip}, V_{\rm gate})$ of Fig.~\ref{Fig2}d, main text.
(a) Experimental branching distance of LL+1. (b) Simulated branching distance of LL+1. (c) Experimental branching distance of LL-2. (d) Simulated branching distance of LL-2.
Insets show the parts of the images in the main text that are used to determine $\Delta x$
with dots that mark the observed maxima in $dI/dV(x_{\rm tip})$ lines, respectively LDOS($x_{\rm tip}$) lines. These maxima are used for distance determination indicated in b.
}
\label{fig:1dmodel}
\end{figure}

\subsection{Interpolations within the Tight Binding Simulations}
\label{sec:InterpolTB}

Poisson simulations are performed for a grid of 22 different $x_{\rm tip}$ and 14 different $V_{\rm gate}$. In the TB simulations we obtain densely sampled plots of the LDOS($x_{\rm tip}$, $V_{\rm gate}$) by employing an interpolation scheme shifting the calculated LDOS below the tip rigidly via a local potential shift. While we use a linear interpolation for each $\Phi_{\mathrm{tot}}(\mathbf{x}=\mathbf{x}_{\rm{tip}})=: \Phi_{\mathrm{tot}}^{\rm tip}$  (\cref{e:tot_potential}) between adjacent $x_{\rm tip}$, we employ a capacitively motivated interpolation along $V_{\rm gate}$ relying on $V_{\rm gate}\kl{\Phi_{\mathrm{tot}}^{\rm tip}} = \zeta \int^{\Phi{\mathrm{tot}}}_{E_{\text{F}}} \text{DOS}(\epsilon) \text{d}\epsilon $ with $\zeta$ describing the inverted capacitance between tip and TIQD. It is used as a fit parameter accounting for the slightly different $v_{\rm F}$ between experiment and TB model.
For calculations of LDOS($x_{\rm tip}$, $V_{\rm sample}$) (Fig.~\ref{Fig4}a, main text, Fig.~\ref{FigS7}a), we employ a rigid energy shift of the LDOS in the center of the TIQD after calculating it for $V_{\rm sample}=0$\,V by $\Phi_{\mathrm{TIQD}}^{(\mathrm{TB})}=0.1\cdot e V_{\mathrm{sample}}$. The lever arm $\beta = 0.1$ is estimated from the Poisson simulations.
The linear shift is justified by the relatively shallow TIQD with $\Phi_{\mathrm{TIQD}}^{(\mathrm{TB})}(\mathbf{x}=
\mathbf{x}_{\rm tip})\leq 20$\,meV. This shallow potential does not enable screening effects originating from different bulk Landau levels at $E_{\rm F}$.\\
 \begin{figure}
\centering
\begin{tikzpicture}
	\node[inner sep=0pt] (russell) at (0,0)   {	\includegraphics[trim= 0 0 0 0, clip,width=0.49\textwidth]{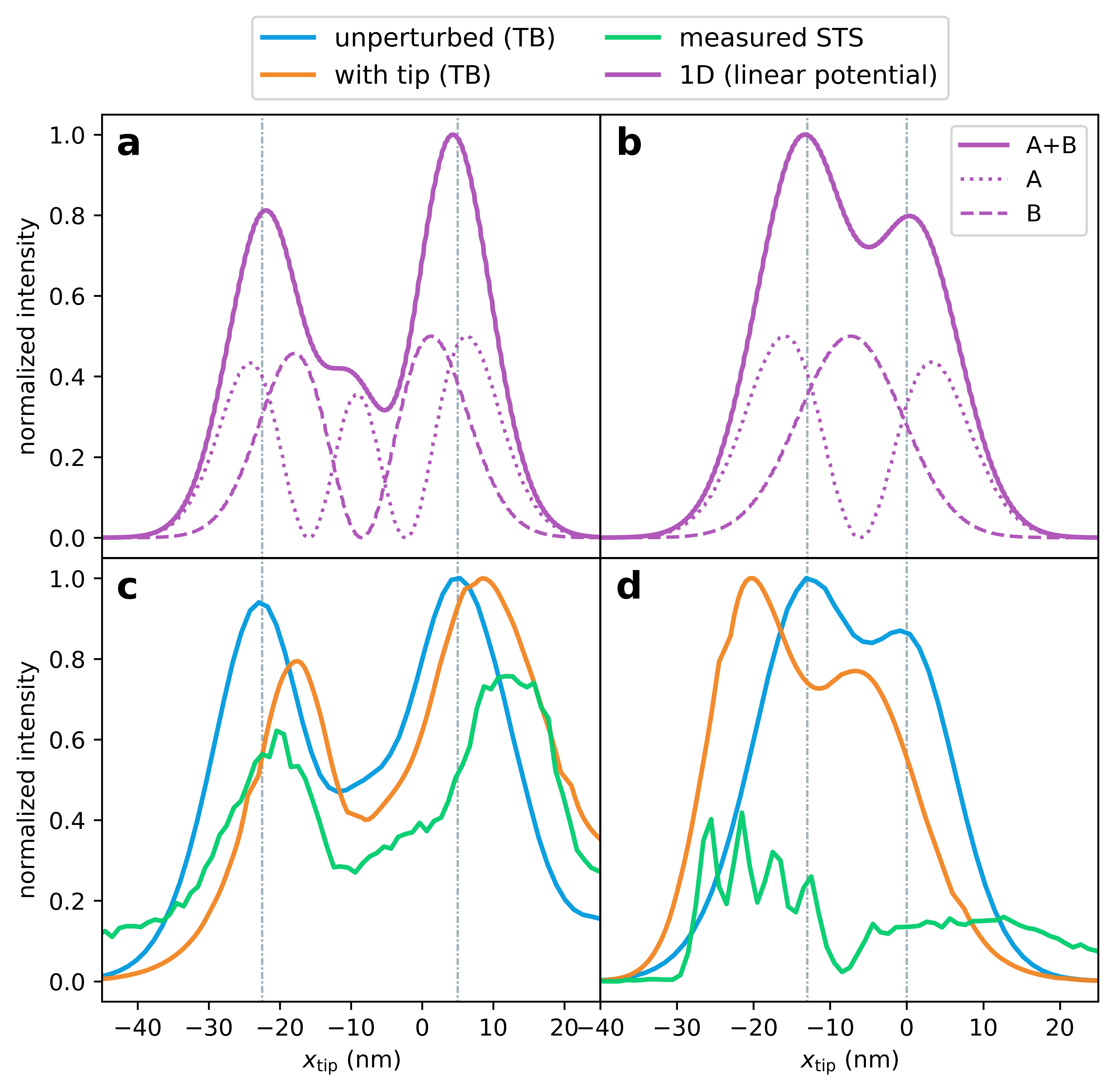}};
\end{tikzpicture}
\caption{
(a) Numerical 1D TB solution of the squared graphene Landau level wave function belonging to LL$\shortminus2$ in a linear potential with slope $-0.8$\,meV/nm (full line). The dashed and dotted curves are the sublattice contributions. (b) Same as subfigure a, but for LL$+1$ at a potential slope of $+0.8$\,meV/nm.
(c,d) Zoom-ins of Fig.~\ref{Fig3}d, main text, for the LL features belonging to LL$\shortminus2$ (c) and LL$+1$ (d). Vertical dashed lines highlight the agreement of distances between maxima.
}
\label{figa:1dmodel}
\end{figure}

\subsection{Additional Comparison between Measured and Simulated Data}
\label{sec:Comparison}

Figure~\ref{fig:1dmodel} shows evaluated lateral distances between the two peaks in the branching features as observed in $dI/dV_{\rm sample}(x_{\rm tip}, V_{\rm gate})$ of Fig.~\ref{Fig1}g, main text, and LDOS$(x_{\rm tip}, V_{\rm gate})$ of Fig.~\ref{Fig2}d, main text. The insets show the relevant areas of the images from the main text. We evaluate the two maxima
of $dI/dV_{\rm sample}(x_{\rm tip})$ and LDOS$(x_{\rm tip})$ at various $V_{\rm gate}$ and determine their mutual distances.
We concentrate on the edge state features that are not (LL-2) or only weakly (LL+1) perturbed by charging lines in the experiment.
Obviously, the experimental distances barely change with $V_{\rm gate}$ as expected for an edge state that originates from the Landau level wave functions within the Landau gauge.\cite{Joynt1984}
The distance is larger for LL-2 as for LL+1 as also expected from the antinodal structure of the corresponding wave functions (Fig.~\ref{figa:1dmodel}a--b). However, the distances found within the tight binding model are smaller by up to 30\,\% and exhibit a weak trend with $V_{\rm gate}$. The latter indicates that Landau level wave functions are mixed by
the (variation in) slope of the potential as the dot moves across the lateral interface.  The origin of the former is unclear, but might be related to electron-electron repulsion that aims to separate electron density maxima and is not included in the tight binding model.

 \begin{figure*}[hpt]
    \centering
    \includegraphics[width=0.9\textwidth]{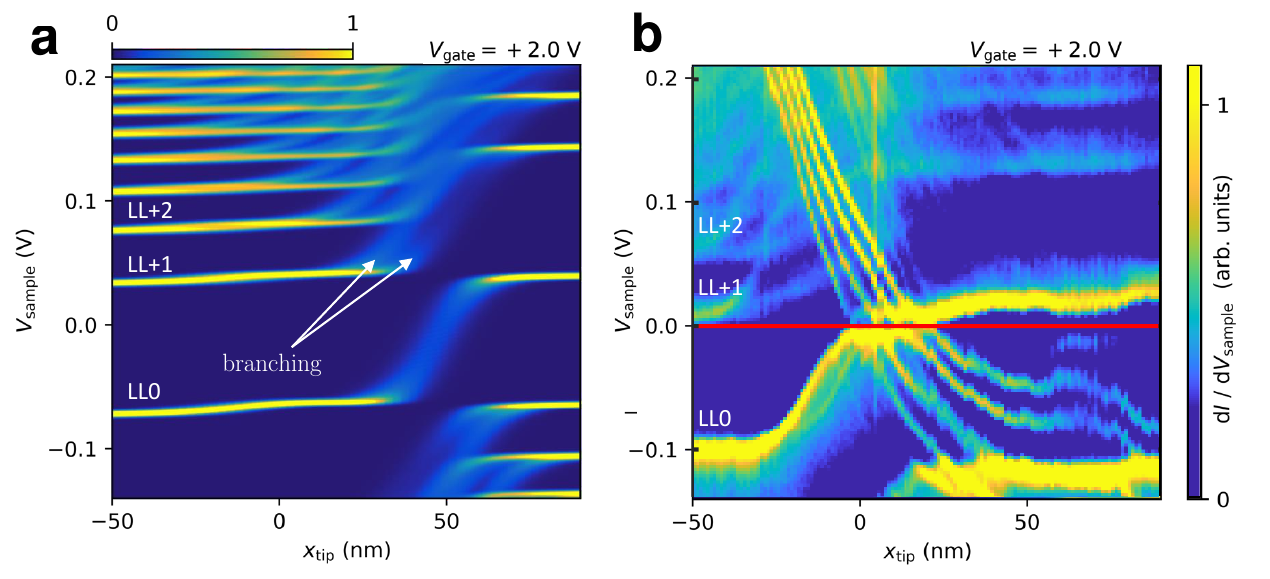}
    \caption{(a) Simulated LDOS($V_{\rm sample}$, $x_{\rm tip}$) across the lateral interface,
        while including the TIQD,
        $V_{\rm{gate}}=2.0$\,V.
        (b) Measured $dI/dV_{\rm sample}(V_{\rm sample}, x_{\rm tip})$, $V_{\rm gate}=2.0$\,V, $I_{\rm stab}=200$\,pA, $V_{\rm stab}=-250$\,mV.
}
    \label{FigS7}
\end{figure*}

Figure~\ref{figa:1dmodel}a--b shows simplified Landau level wave functions of graphene within a linear potential as calculated by a one-dimensional tight binding model. The slope of the potential ($0.8$\,meV/nm) is chosen in between the slopes observed on the plateaus at the interface within the Poisson simulations ($0.1-0.2$\,meV/nm) and the average slope found across the lateral interface ($1.5-3$\,meV/nm) (Fig.~\ref{Fig2}c, main text).
The additionally displayed two sublattice contributions reveal the well-known one- and two-fold antinodal structure for LL+1 as well as the two- and three-fold antinodal structure for LL-2, representing the chiral symmetry of graphene in analogy to the quantum dot solutions of eqs.~(\ref{valley}). However, remarkably, the resulting peaks are different in height showing that the wave functions are not pure Landau gauge solutions, but the solutions are mixed LL wave functions due to the influence of the potential slope. The simplified model nicely reproduces peak distances and relative peak heights of the more complex tight binding simulations that include the detailed potential of the Poisson solver (blue lines in Fig.~\ref{figa:1dmodel}c--d) as well as the ones that additionally consider the TIQD (orange lines in Fig.~\ref{figa:1dmodel}c--d). They also reproduce the trend of different peak heights as found in the experiment, but underestimate the experimental peak distances.

We conclude that the observed edge states are largely given by the Landau gauge wave functions of the two sublattices. However, the interface potential gradient leads to a small LL wave function mixing, that implies a slightly larger (smaller) wave function peak located at the more attractive (repulsive) potential side of the interface. Moreover, it is likely that electron-electron repulsion increases the inter-peak distance again via Landau level mixing.
We checked with the 1D tight binding model that potential slopes up to $3$\,meV/nm do not change the inter-peak distance by more than 1.5\,nm and, thus, cannot explain the observed larger distances in the experiment.

Finally, we provide additional data corroborating the good, semi-quantitative
agreement between our experiment and the TB simulations including the TIQD.
Besides the comparison between experimental data and TB simulations in the main text (Fig.~\ref{Fig1}g, \ref{Fig2}d, \ref{Fig4}a--d), a
comparison of measured
$dI/dV_{\rm sample}(V_{\rm sample}, x_{\rm tip})$   and calculated LDOS($V_{\rm sample}$, $x_{\rm tip}$) at positive $V_{\rm gate}$ is shown in Fig.~\ref{FigS7}. As always, the parameters determined in section~S2--S3 are used as base providing the potentials for the TB simulations. The positive $V_{\rm gate}$ results in a p-n interface with $\nu=2$ on the left and $\nu =-2$ on the right of the interface. Again, the general agreement between experimental data and simulations is very good with the exception of the additional charging lines in the experimental data. Most importantly, the branching of LL$n$ features is again observed in the experiment and in the simulation, here very pronounced for the $n>0$ LLs.

\subsection{Analysis of $dI/dV_{\rm sample}$ ($V_{\rm gate}$, $V_{\rm sample}$) at a Fixed Lateral Position}
\label{sec:fixedlateral}

\begin{figure*}
    \centering
    \includegraphics[width=0.999\textwidth]{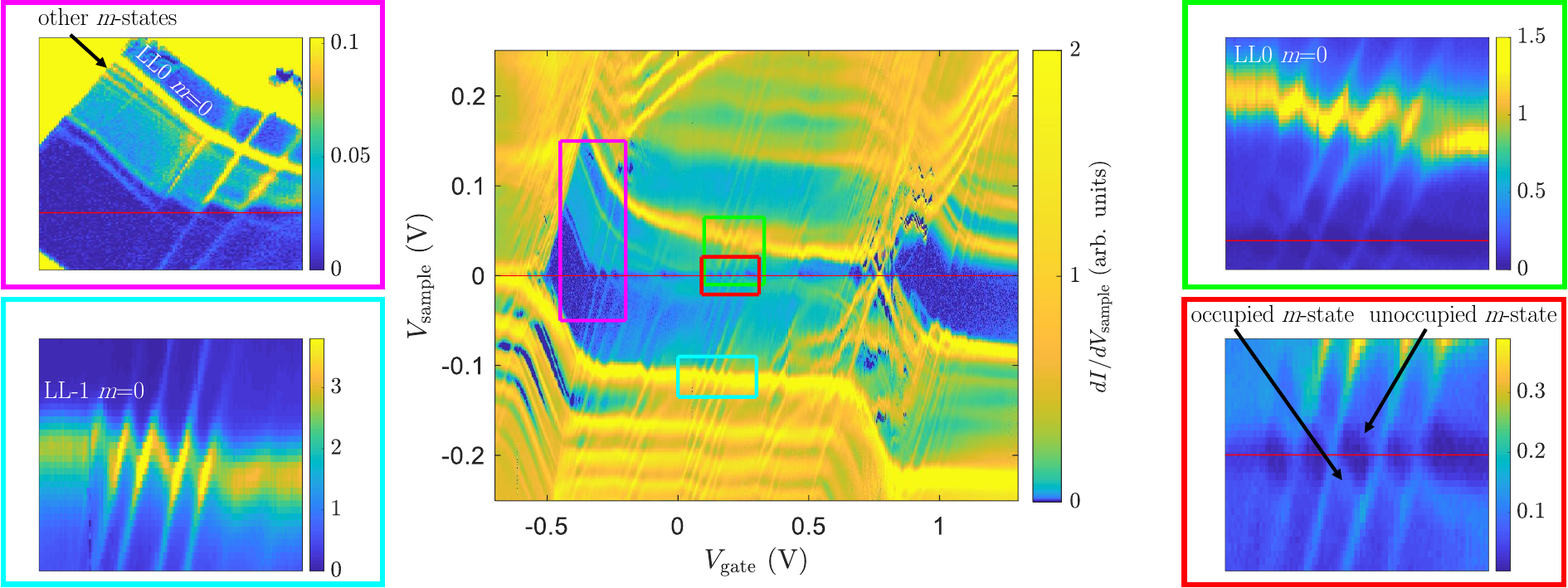}
    \caption{Zooms into $dI/dV_{\rm sample}$ ($V_{\rm gate}$, $V_{\rm sample}$) around the LL0 plateau at $E_{\rm F}$ at fixed position $x_{\rm tip} \ll 0$\,nm (same data as Fig.~\ref{Fig1}b, main text), $I_{\rm stab}=1$\,nA, $V_{\rm stab}=-250$\,mV. The areas of the four zooms are marked in the central image by a frame of the same color. Magenta frame: LDOS lines of the various $m$ states that belong to LL0. Cyan frame: Kinks in the LDOS line belonging to LL-1 that appear each time when a charging line is crossing. Green frame: same as cyan frame for the LDOS line belonging to the $m=0$ state of LL0. Red frame: Coulomb diamonds at $E_{\rm F}$ belonging to a higher $m$ state of LL0. The occupied and unoccupied version of the same $m$ state is marked.
}
    \label{FigS6a}
\end{figure*}

Figure~\ref{FigS6a} shows several zooms into the map of $dI/dV_{\rm sample}$ ($V_{\rm gate}$, $V_{\rm sample}$) as displayed in Fig.~\ref{Fig1}b, main text, that is recorded at a position far away from the lateral interface.
In the upper left corner, the zoom showcases the appearance of several $m$ states of LL0 that are confined at different energies within the TIQD. Obviously, these lines largely run in parallel along $V_{\rm gate}$ stressing a similar energy change by the back gate voltage for all of these TIQD states. As discussed in the main text, the brightest line belongs to $m=0$. It is found at largest $V_{\rm sample}$ rendering the TIQD hole-type. i. e. higher $m$ states are at lower energy. Coulomb diamonds appear when the different $m$ states cross $E_{\rm F}$ (red line).
The zooms in the lower left and the upper right of Fig.~\ref{FigS6a}
feature the kinks in the LDOS lines of $m=0$ states away from $E_{\rm F}$ ($V_{\rm sample}=0$\,V) that appear whenever a charging line is crossing. As described in the main text, this showcases the Coulomb staircase effect, i.e. the LDOS is shifted by the Coulomb repulsion of the additional charge within the TIQD.\cite{Freitag2016} The upper right zoom, moreover, features a quadruplet of rather equidistant charging lines. The four rightmost ones have a similar mutual distance, while the fifth one exhibits a larger distance to the fourth one. This fourfold bunching is caused by the fourfold spin and valley degeneracy of each $m$ state in graphene. By following the charging lines down to $E_{\rm F}$ (red line) and comparison with the central image, it is also apparent that these charging lines mark the charging of a higher $m$ state of LL0.

A zoom into  the crossing area of these charging lines with $E_{\rm F}$ (lower right zoom) reveals the so-called Coulomb diamonds rather clearly. They result from the simultaneous crossing of the LDOS features of the $m$-states and the charging lines across $E_{\rm F}$. Naturally, the subsequent charging of a single $m$-state must imply the simultaneous presence of unoccupied and occupied versions of the $m$-state, except after filling of the fourth degeneracy level. This is nicely visible as LDOS lines propagating in parallel  above and below $E_{\rm F}$ (arrows). The pair of LDOS lines is separated by the charging voltage that must be provided by the tip to place one more electron into the TIQD. The two state energies increase in parallel for more negative $V_{\rm gate}$ and jump back down if an additional hole is charged into the TIQD, i.e., if a charging line crosses. After four such jumps, the $m$-state is completely empty and the next $m$-state moves towards $E_{\rm F}$ for charging. Hence, again the four visible Coulomb diamonds in the lower right zoom of Fig.~\ref{FigS6a} indicate the fourfold degeneracy of the corresponding $m$ state in graphene.\\

\begin{figure}
    \centering
    \includegraphics[scale=0.59]{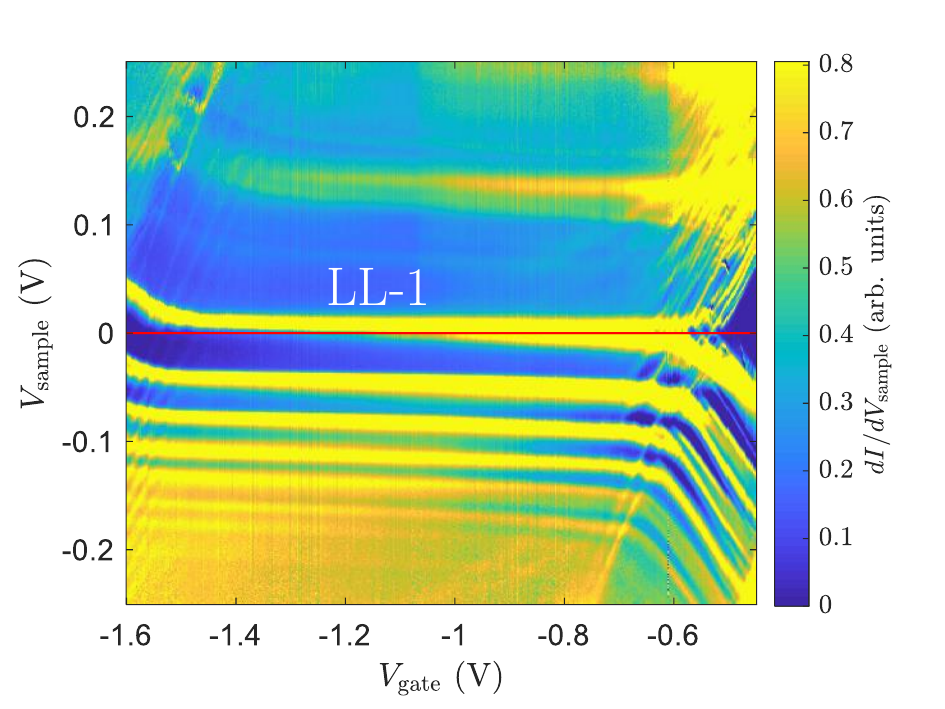}
    \caption{$dI/dV_{\rm sample}$ ($V_{\rm gate}$, $V_{\rm sample}$) (zoom into Fig.~\ref{Fig1}a, main text), $x_{\rm tip} \ll 0$\,nm, $I_{\rm stab}=1$\,nA, $V_{\rm stab}=-250$\,mV. The LL-1 plateau and its charging lines starting from the right end of the plateau are visible.
}
    \label{FigS6b}
\end{figure}

Figure~\ref{FigS6b} features the plateau at $E_{\rm F}$ of the LDOS line belonging to LL-1. The most bright charging lines appear on the right end of the plateau followed by weaker charging lines towards the left. As explained in the main text, this supports our classification of the TIQD as a hole-type dot. The ($m=0$)-state is the one with the highest probability density in the center of the quantum dot and, hence, leads to the strongest charging line by its strongest Coulomb repulsion acting on the states that are probed by the tip. The fact that this ($m=0$)-state is charged at the largest $V_{\rm gate}$ further corroborates the assignment of the TIQD to a hole-type band bending.

Notice that additional bright charging lines appear in the upper left corner of Fig.~\ref{FigS6b}. They are likely caused by the charging of the ($m=0$)-state of LL-2.

\subsection{Branching of Landau Levels at Different $V_{\rm gate}$}
\label{sec:Branching}

\begin{figure*}
    \centering
    \includegraphics[width=0.9\textwidth]{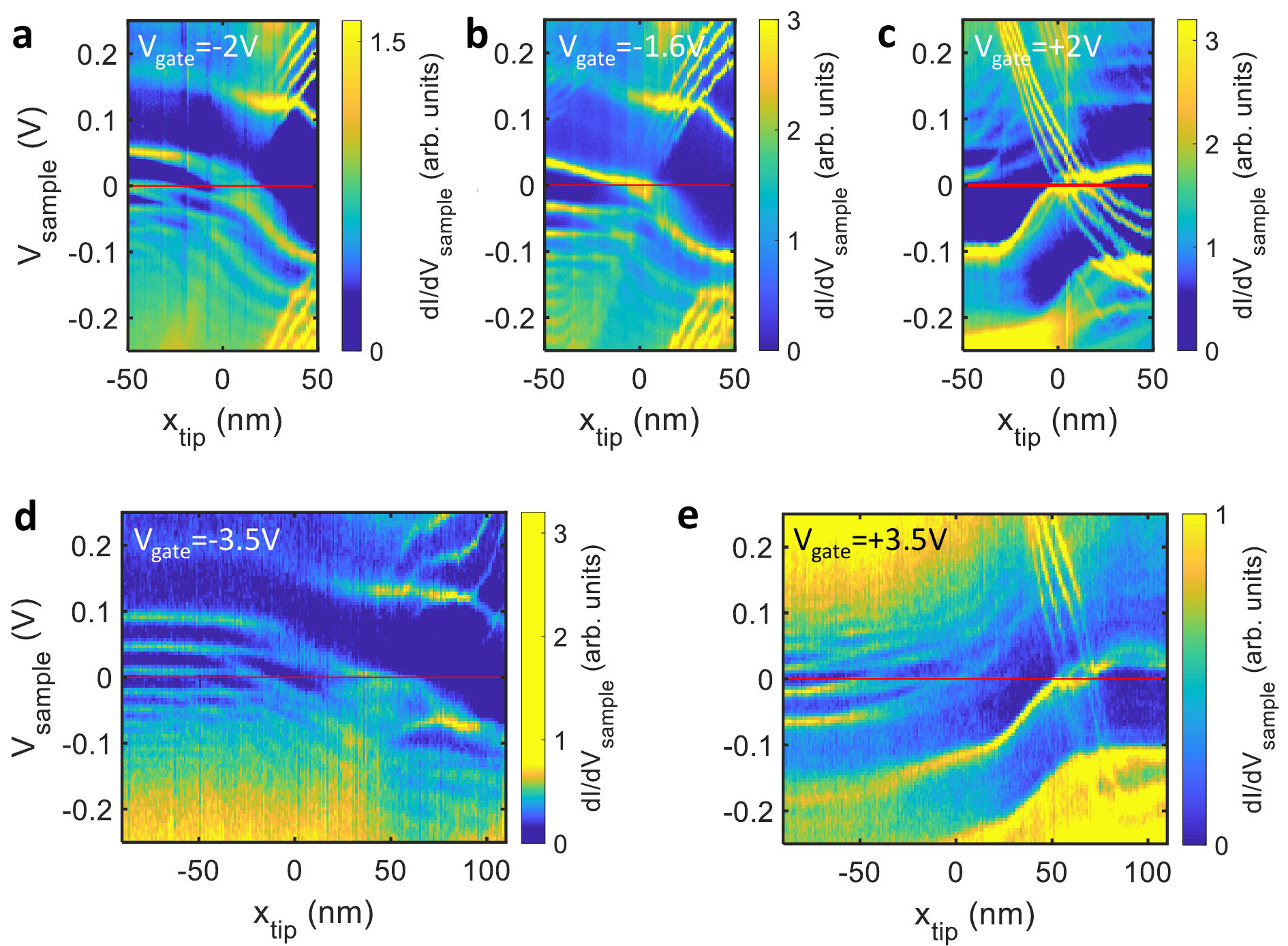}
    \vspace{-0.3cm}
    \caption{Comparison of $dI/dV_{\rm sample}( x_{\rm tip},\,V_{\rm sample})$ across the lateral interface for several $V_{\rm gate}$ as marked.
    (a) $V_{\rm gate}=-2.0$\,V, $I_{\rm stab}=200$\,pA, $V_{\rm stab}=-250$\,mV (same as Fig.~\ref{Fig4}b, main text).
    (b) $V_{\rm gate}=-1.6$\,V, $I_{\rm stab}=200$\,pA, $V_{\rm stab}=-250$\,mV.
    (c) $V_{\rm gate}=+2.0$\,V, $I_{\rm stab}=200$\,pA, $V_{\rm stab}=-250$\,mV (same as Fig.~\ref{FigS7}b)
    (d) $V_{\rm gate}=-3.5$\,V, $I_{\rm stab}=200$\,pA,  $V_{\rm stab}=-500$\,mV (same as Fig.~\ref{Fig1}f, main text) .
    (e) $V_{\rm gate}=+3.5$\,V, $I_{\rm stab}=200$\,pA,  $V_{\rm stab}=-500$\,mV.
    }

    \label{Fig_comp_split}
\end{figure*}

Figure~\ref{Fig_comp_split} shows the $dI/dV_{\rm sample}( x_{\rm tip},\,V_{\rm sample})$ maps across the lateral interface for various $V_{\rm gate}$ applied to the left side of the map areas. The changing filling factor on the left is visible by the number of Landau levels below $E_{\rm F}$ (red line). The filling factor $\nu$  on the very left changes from $\nu=-14$ at $V_{\rm gate}=-3.5$\,V, i.e. the LL-3 is  not occupied with electrons, but LL-4 is occupied, to $\nu=10$ at $V_{\rm gate}=3.5$\,V, i.e. LL2 is occupied, but LL3 is not. Moreover, the branching of the LL features appears at all $V_{\rm gate}$. At negative $V_{\rm gate}$, the branching looks very similar for different $V_{\rm gate}$ except that the lateral onset of branching is shifted. In contrast, at $V_{\rm gate}=+2$\,V, the branching appears to be stronger leading to intersections of different branching lines as in the TB calculations (Fig.~\ref{FigS7}). In that case, charging lines directly cross the branching areas indicating a strong influence of the TIQD. In turn, in case of weak interference from the TIQD, the branching is rather stable supporting the interpretation that it is caused by the antinodal structure of the edge states at the interface. A more detailed investigation of the strengths of the various branching strengths of occupied and unoccupied LLs as a function of $V_{\rm gate}$ is beyond the scope of this study.

Finally, one nicely sees that the LLs mostly exhibit plateaus, if they cross $E_{\rm F}$ indicating the development of compressible stripes at the interface.\cite{Chklovskii1992,Lier1994}

\subsection{Influence of Strain on Branching}
\label{sec:strain}
\begin{figure*}
    \centering
    \includegraphics[width=0.8\textwidth]{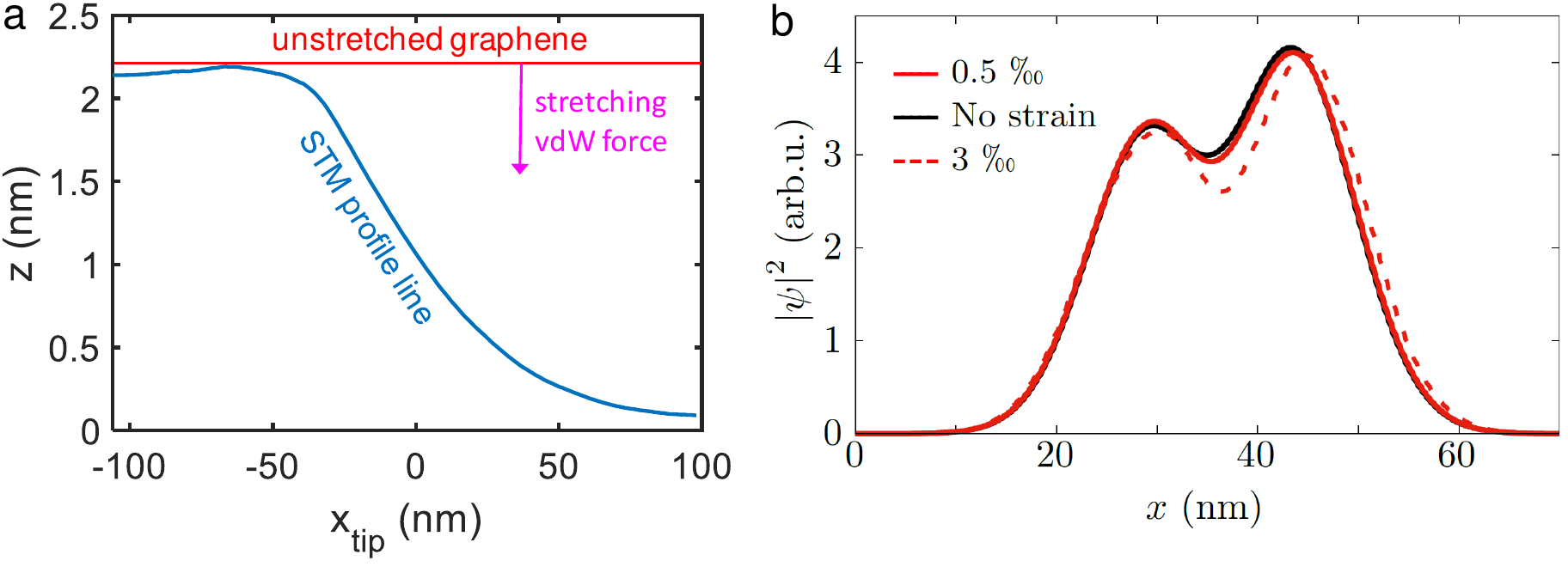}
    \vspace{-0.5cm}
    \caption{ (a) Blue line: profile line across the step edge due to the graphite gate as measured by STM (see also Fig.~\ref{Fig1}c--d). Red line: tentative position of relaxed graphene directly prior to contact with the hBN. Pink arrow: Force that pulls the graphene downwards to the hBN. (b) Wave function corresponding to LL1 determined by a tight binding calculation without strain (black), with a Gaussian strain profile of amplitude 0.05 \% and FWHM 60 nm (red full line), and with amplitude 0.3 \% and FWHM 60 nm (red dashed line).}
    \label{FigS20}
\end{figure*}

One might wonder, if the presence of the step edge visible in Fig.~\ref{Fig1}d leads to strain that eventually causes the branching of the LL features in $dI/dV_{\rm sample}(x_{\rm tip})$. To exclude such a scenario, we estimate the strain in the following. The step edge visible in Fig.~\ref{Fig1}d has a height of 2.1\,nm and a width of $\sim 70$\,nm according to its line profile (Fig.~\ref{FigS20}a). The line profile exhibits a continuous curvature with nearly Gaussian shape across the edge. The smooth shape suggests a direct contact of the graphene to the underlying hBN. The graphene is deposited in a separate step after the hBN, such that the hBN already covers the graphite edge prior to graphene transfer. Hence, there is no obvious reason that the graphene should be particularly stretched at the step edge. During transfer, the graphene just sees a minimally bended hBN below. But even if one assumes that the graphene profile develops from a relaxed, initially flat graphene exactly parallel to the SiO$_2$ substrate (red line in Fig.~\ref{FigS20}a), the resulting strain from stretching it to the measured profile line would be below ~0.05\,\% only.
This is roughly the same magnitude as the typical strain fluctuations for graphene samples on flat hBN that exhibit a rms value of 0.05\,\% as well.\cite{Couto2014,Neumann2015} Hence, if strain of this small magnitude would cause a peak splitting, such a splitting would appear everywhere, not only at the step edge, in clear contrast to the experiment.

To quantitatively assess the influence of strain on the LL wave functions we consider a strain of 0.05\,\% as a maximum of a Gaussian profile with full width at half maximum (FWHM) of 60\,nm. We modify the hopping parameters accordingly in the TB simulation. We find only minimal changes in the two component Landau level wave function (see Fig.~\ref{FigS20}b, black vs. red line). The double peak structure barely changes due to this strain. To asses the effect of even larger strain, we increased the strain in the calculation by a factor of six and still found only minor qualitative changes (dashed line, Fig.~\ref{FigS20}b). For the 1D step edge, we only expect a strain gradient perpendicular to the edge, and thus no pseudomagnetic field that requires a two-dimensional strain distribution.\cite{Guinea2009} However, even if one assumes a circular symmetric Gaussian bump of the same profile as the step, the pseudomagnetic field would be ~200 mT only,\cite{Guinea2009,Georgi2017} much smaller than the externally applied magnetic field (7\,T). The difference in Landau quantization due to such a small pseudomagnetic field would result in an energy splitting between the two Dirac cones of $\sim 2$\, meV.\cite{Guinea2009} This unrealistic strain scenario, thus, would still be significantly too small to explain the observed splittings during branching of about 25\,meV.

Consequently, we can safely exclude that strain is a major factor for the observed branching of LDOS features at the lateral interface.

\end{document}